\documentclass[sigconf]{acmart}
\AtBeginDocument{%
  }

\setcopyright{acmlicensed}
\copyrightyear{2018}
\acmYear{2018}
\acmDOI{XXXXXXX.XXXXXXX}
\acmConference[Conference acronym 'XX]{Make sure to enter the correct
  conference title from your rights confirmation email}{June 03--05,
  2018}{Woodstock, NY}
\acmISBN{978-1-4503-XXXX-X/2018/06}

\usepackage{booktabs}   
\usepackage{subcaption} 
\usepackage{tikz}
\usepackage{amsmath}
\usepackage{subcaption}
\usepackage{pifont}
\usepackage{booktabs}
\usepackage{graphicx}
\usepackage{array}
\usepackage{fontawesome}
\usepackage{multirow}
\usepackage{makecell}
\usepackage{array}
\usepackage{graphicx}
\usepackage{algorithm}
\usepackage{algpseudocode}
\usepackage{fontawesome}

\newcommand{\squishlist}{
  \begin{list}{$\bullet$}
    { \setlength{\itemsep}{1pt}
      \setlength{\parsep}{1pt}
      \setlength{\topsep}{2.5pt}
      \setlength{\partopsep}{0.5pt}
      \setlength{\leftmargin}{1em}
      \setlength{\labelwidth}{1em}
      \setlength{\labelsep}{0.6em}
    }
  }
  \newcommand{\squishend}{
  \end{list}
}

\begin{document}

\title{KBest: Efficient Vector Search on Kunpeng CPU}



\author{Kaihao Ma$^{\dag}$,  Meiling Wang$^{\dag}$, Senkevich Oleg$^{\dag}$, Zijian Li$^{\dag}$, Daihao Xue$^{\dag}$, Malyshev Dmitriy$^{\P}$, Yangming Lv$^{\dag}$, Shihai Xiao$^{\dag}$, Xiao Yan$^{\ddag}$, Radionov Alexander$^{\dag}$, Weidi Zeng$^{\dag}$,   Yuanzhan Gao$^{\dag}$, Zhiyu Zou$^{\dag}$,   Xin Yao$^{\dag}$,  Lin Liu$^{\dag}$, Junhao Wu$^{\dag}$, Yiding Liu$^{\dag}$, Yaoyao Fu$^{\dag}$,   Gongyi Wang$^{\dag}$, Gong Zhang$^{\dag}$, Fei Yi$^{\dag}$, Yingfan Liu$^{\S}$} 
\affiliation{
    \institution{$^\dag$Huawei Technologies Ltd. \quad $^\ddag$ Wuhan University  \quad $^\P$ Higher School of Economics \quad $^\S$ Xidian University}
    \country{}
}

\renewcommand{\shortauthors}{Ma et al.}

\newcommand{\redbold}[1]{\textbf{\color{red}#1}}
\newcommand{\xiao}[1]{\textcolor{red}{Xiao:#1}}
\newcommand{\name}{{\textsf{KBest}}}

\algtext*{EndWhile} 
\algtext*{EndIf}    
\algtext*{EndFor}   

\newcommand{\stitle}[1]{\vspace*{0.4em}\noindent{\bf #1.\/}}

\begin{abstract}
    Vector search, which returns the vectors most similar to a given query vector from a large vector dataset, underlies many important applications such as search, recommendation, and LLMs. To be economic, vector search needs to be efficient to reduce the resources required by a given query workload. However, existing vector search libraries (e.g., Faiss and DiskANN) are optimized for x86 CPU architectures (i.e., Intel and AMD CPUs) while Huawei Kunpeng CPUs are based on the ARM architecture and competitive in compute power. In this paper, we present \textsf{KBest} as a vector search library tailored for the latest Kunpeng 920 CPUs. To be efficient, \textsf{KBest} incorporates extensive hardware-aware and algorithmic optimizations, which include single-instruction-multiple-data (SIMD) accelerated distance computation, data prefetch, index refinement, early termination, and vector quantization. Experiment results show that \textsf{KBest} outperforms SOTA vector search libraries running on x86 CPUs, and our optimizations can improve the query throughput by over 2x. Currently, \textsf{KBest} serves applications from both our internal business and external enterprise clients with tens of millions of queries on a daily basis.
\end{abstract}

\maketitle

\section{Introduction}\label{sec:intro}

\begin{figure}[!t]
	\centering
	\includegraphics[width=\columnwidth]{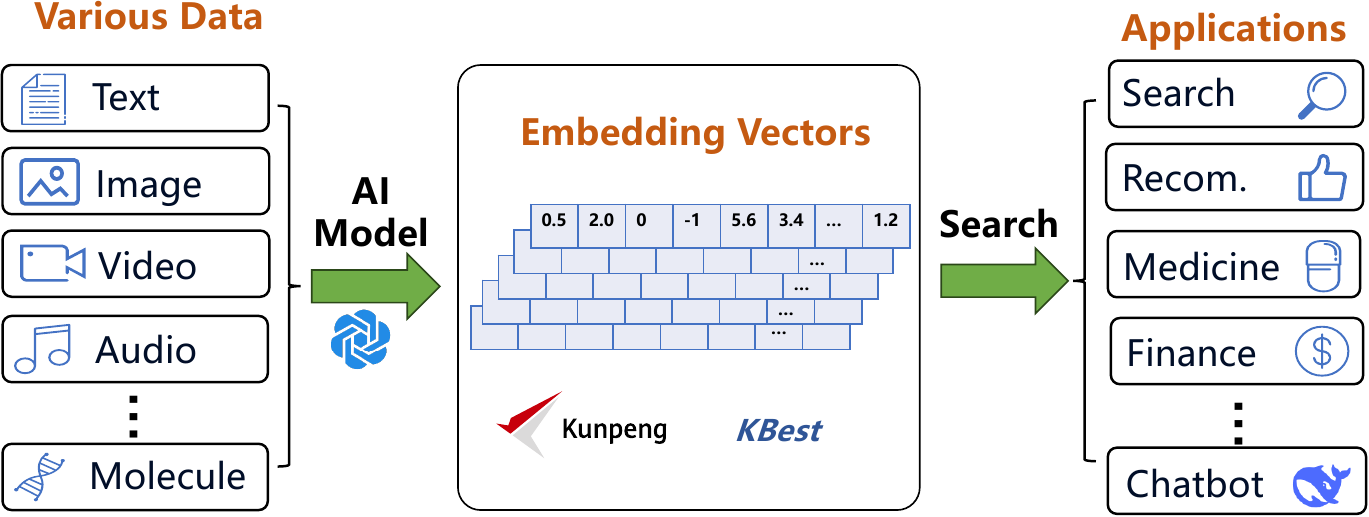}
	\caption{Vector embeddings and applications of vector search}
	\label{fig:anns-system}
	\vspace{-3mm}
\end{figure}

With the development of machine learning, many embedding models~\cite{bge,genept,learning,zhaolearning} are proposed to map data objects (e.g., texts, images, videos, and molecules) to vectors that encode their semantics. On these embedding vectors, similarity is a key notion. For instance, two images look similar if their embeddings are similar (e.g., as measured by Euclidean distance), a text description matches a video if their have similar embeddings, and a user may like a product if their embeddings are similar. Therefore, vector search, which returns the vectors most similar to a given query vector from a vector dataset, is a basic operation on embeddings~\cite{anns}. As shown in Figure~\ref{fig:anns-system}, vector search underlies many important applications such as content search (e.g., for images and videos)~\cite{pq,localpq,visual}, recommendation (e.g., for e-commerce or contents)~\cite{spann,fbembedding,anna,van2016learning}, medicine~\cite{medicine1,medicine2,medicine3}, finance~\cite{finance1}, and even LLM chat-bots~\cite{llmbot1,llmbot2}. Vector search usually needs to handle large vector datasets (e.g., with millions or even trillions of high-dimension vectors)~\cite{largescale} and meet stringent performance requirements (e.g., returning  search results within 10ms and serving millions of queries per second)~\cite{vectorsearchsurvey}. As such, vector search should be efficient (i.e., achieving a high query throughput) to reduce the required resources for serving a given workload and thus monetary cost.

Due to the importance of efficiency, there are several highly optimized libraries for vector search. For instance, HNSWlib~\cite{hnswlib} provides an efficient implementation for HNSW~\cite{hnsw}, which is a proximity graph index for vector search and shown to achieve good performance. Developed by Meta, FAISS~\cite{faiss} supports both inverted file (IVF)~\cite{pq} and proximity graph as the indexes and allows to use vector quantization for efficient distance computation. DiskANN~\cite{diskann} is proposed by Microsoft and supports both memory-based and disk-based vector search with a novel proximity graph index called Vamana. Powered by Google, ScaNN~\cite{scann} uses registers to compute vector distance efficiently via table lookup. 

However, these vector search libraries target x86 (i.e., Intel and AMD) CPUs while ARM CPUs are becoming competitive. In particular, Huawei started the ARM-based Kunpeng CPU series~\cite{HiSilicon,hikunpeng2023} in 2013 and the production lines maintains an annual capacity of hundreds of thousands of CPUs. With 80 cores at 2.9GHz, the latest Kunpeng 920 CPU matches AMD 9654 (with 96 cores and 2.4GHz) in compute power. Currently, Kunpeng 920 CPUs are widely deployed for both our internal business and external cloud services. As such, it is crucial to develop an efficient vector search library for Kunpeng CPUs in order to support the applications that rely on vector search. However, developing such a library is challenging because (i) ARM CPUs have different hardware characteristics and instruction sets from x86 CPUs, and thus a deep understanding of ARM CPUs and extensive engineering efforts are required; (ii) vector search has a long research history, and thus an extensive survey is required to understand the best practices and integrate them into our library.


In this paper, we present \name{} (Kunpeng Blazing-fast embedding similarity search thruster) as an efficient vector search library tailored for ARM-based Huawei Kunpeng CPUs. \name{} targets memory-based vector search, which assumes that the vectors fit in the main memory of a machine and is the most common scenario, and adopts proximity graph indexes, which achieve the best performance for vector search. To improve efficiency, \name{} incorporates comprehensive hardware-aware and algorithmic optimizations. In particular, we leverage the single-instruction-multiple-data (SIMD) instructions of ARM to implement efficient vector distance computation, which is the basic operation in vector search, and conducts software prefetch to reduce the cache miss when accessing vectors following the edges of proximity graph indexes. We also use huge memory pages and align the vectors with cache lines to reduce memory management overheads. From the algorithm perspective, we introduce a refinement step, which checks the 2-hop neighbors in an existing proximity graph index to improve its quality, and propose a lightweight graph reordering algorithm to renumber the vectors for improved data access locality. Besides, we also design a method to terminate the processing of a query early when its neighbors have been identified and allow users to flexibly configure their vector quantization algorithms, which reduce distance computation complexity by using compressed vectors and are crucial for efficiency.


We evaluate \name{} on 4 real-world vector datasets and compare with 3 SOTA x86-based vector search libraries. The results show that \name{} running on Kunpeng 920 CPU outperforms existing vector search libraries running on AMD 9654 CPU, highlighting the huge performance potential of ARM-based platforms. Ablation studies also suggest that our optimizations are effective by improving the query throughput by over 2x. Currently, \name{} is widely used from both our internal business and external users, serving tens of millions of queries in a daily basis.

To summarize, we make the following contributions:
\squishlist

\item We design and implement \name{} as the first efficient vector search library for ARM CPUs, incorporating both ARM-specific hardware optimizations and general algorithmic improvements.

\item We design user-friendly API to allows users to easily use \name{} and integrate with existing vector databases.

\item Our optimizations encompass the best practices for CPU-based vector search and can guide followup works.

\squishend

\section{Preliminaries}\label{sec:background}

\begin{figure}[!t]
	\centering
	\includegraphics[width=0.8\columnwidth]{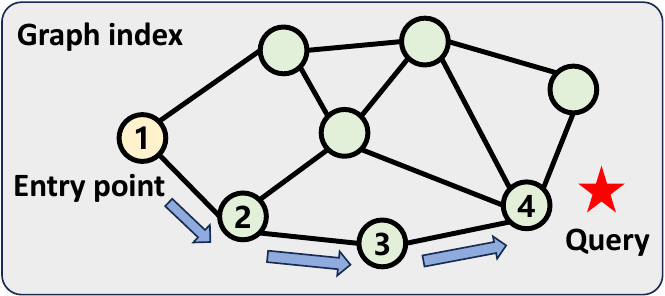}
	\caption{An illustration of vector search on proximity graph indexes, which traverses the graph to identify neighbors}
	\label{fig:graph-index}
	\vspace{-7mm}
\end{figure}

In this part, we introduce the basics of vector search and Huawei Kunpeng CPUs to facilitate subsequent discussions.

\subsection{Vector Search and Proximity Graph Index}

Vector search, also known as nearest neighbor search (NNS), is usually defines as follows.
\begin{definition}
    Given a query vector $\mathbf{q} \in \mathbb{R}^d$ and a  vector dataset $\mathcal{X} = \{\mathbf{x}_1, \mathbf{x}_2, \cdots, \mathbf{x}_n\} \subset \mathbb{R}^d$, find the set $\mathcal{N}_q\subset \mathcal{X}$ of the top-$k$ nearest neighbors for $\mathbf{q}$ such that 
\[
|\mathcal{N}_q|=k \ \text{and} \ \operatorname{dist}(\mathbf{q}, \mathbf{x})\le \operatorname{dist}(\mathbf{q}, \mathbf{x}')  \ \forall \mathbf{x}\in \mathcal{N}_q, \mathbf{x}' \in \mathcal{X}-\mathcal{N}_q,
\]
\end{definition}
Here, $\operatorname{dist}(\mathbf{q}, \mathbf{x})$ is a distance function, such as Euclidean distance (i.e., $\|\mathbf{q} - \mathbf{x}\|_2$) or negative inner product (i.e, $-\langle \mathbf{q}, \mathbf{x} \rangle$). As the dimension of embedding vectors are usually high (e.g., at hundreds), exact NNS requires a linear scan due to the curse of dimensionality~\cite{dimcurse}. To trade for efficiency, approximate NNS (ANNS) is usually used in practice, which returns most rather than all of the top-$k$ nearest neighbors for each query. The quality of an approximate result set $\mathcal{N}_q'$ is typically measured by \textit{recall}, which is defined as $|\mathcal{N}_q' \cap \mathcal{N}_q| / k$. Applications usually require a high recall (e.g., 0.9 or 0.95) for good result quality and a low query latency (e.g., <10ms) for good QoS. The performance of vector search is commonly measured by the QPS (query processing throughput) at specific recall levels. 

\begin{algorithm}[!t]
\caption{Graph Traversal for Vector Search}
\begin{algorithmic}[1]
\State \textbf{Input:} Graph $G$, query $q$, result count $k$, queue size $L$
\State \textbf{Output:} $k$ similar vectors to $q$
\State Initialize a size-$L$ priority queue $Q$ with $(v_1, \Vert v_1-q\Vert)$
\While{$Q$ has unvisited node}
    \State Read the most similar but unvisited node $v$ in $Q$
    \For{each neighbor $u$ of $v$ in $G$}
        \If{distance $\Vert q-u \Vert$ is not computed}
        \State Compute $\Vert q-u \Vert$
        \State Try to insert $(u, \Vert q-u \Vert)$ into $Q$ 
        \EndIf
    \EndFor
\EndWhile
\State \Return The $k$ vectors with the smallest distances in $Q$ 
\end{algorithmic}
\label{alg:beam-search}
\end{algorithm}

Many algorithms and indexes have been designed for vector search including locality sensitive hashing (LSH)~\cite{lsh}, tree-based data structures~\cite{kdtree,rtree}, inverted file (IVF)~\cite{pq}, and proximity graph~\cite{approximate_graph}. Proximity graph is the most efficient for high dimension embedding vectors in that it requires the fewest distance computations to reach the same recall. As shown in Figure~\ref{fig:graph-index}, proximity graph organizes the vector dataset as a graph, where the nodes are vectors and edges connects similar vectors. The number of neighbors for each vector is usually limited by a small number $M$ (e.g., 64). Vector search is conducted by the graph traversal procedure in algorithm~\ref{alg:beam-search}. \textit{candidate queue} $Q$ is a minimum priority queue to manage the distances and return the unvisited node with the smallest distance. Search starts with a random or fixed entry node (i.e., $v_1$) and checks the most similar but unvisited node $v$ by computing distances for $v$'s neighbors. Vector search terminates when $Q$ can no longer be updated and the finally line 10 returns the top-k results with small distances. A larger queue size $L$ improves recall buy computing more distances but consumes longer search time. There are many variants of proximity graph, e.g., HNSW~\cite{hnsw}, NSG~\cite{nsg}, Vamana~\cite{diskann}, SSG~\cite{ssg}, each with different edge selection rules during index building. They usually perform well for different datasets.

\subsection{Huawei Kunpeng CPUs}

\begin{table}[!t]
\centering
\caption{Hardware specifications for Huawei Kunpeng 920 CPU and representative x86 CPUs}
\label{tab:cpu-specification}
\begin{tabular}{cccc}
\hline
\multicolumn{1}{l}{} & \textbf{Intel 8558p} & \textbf{AMD 9654}  & \textbf{Kunpeng 920} \\ \hline
\textbf{Cores}        & 48C     & 96C  & 80C          \\ \hline
\textbf{Threads}        & 96T     & 192T  & 160T          \\ \hline
\textbf{Frequency}       & 2.7GHz      & 2.4 GHz   & 2.9GHz            \\ \hline
\end{tabular}
\end{table}

Huawei's Kunpeng CPU series are based on the ARM architecture and have more than ten years of development history. The architectural evolution spans from the first-generation Hi1610 in 2013 to the current fourth-generation Kunpeng 920 in 2019, fabricated using 7nm process technology. As shown in Table~\ref{tab:cpu-specification}, Kunpeng 920 CPU processors features up to 80-core configuration with 2.9GHz clock frequency, demonstrating competitive performance against SOTA X86 platform Intel Xeon and AMD EPYC processors. Currently Kunpeng 920 CPU processors have been widely adopted in many fields such as cloud computing infrastructures and database systems.

\stitle{Instruction set of ARM CPUs} ARM-based Kunpeng CPUs and x86 processors exhibit similar functionalities but in different forms. For SIMD acceleration, Kunpeng leverages 128-bit NEON (ARM Advanced SIMD) and scalable bit-length SVE (Scalable Vector Extension) instructions, offering adaptive vectorization for irregular dimensions, contrasting with 256-bit AVX2 (Advanced Vector Extensions 2) and 512-bit AVX-512 (Advanced Vector Extensions 512) on x86 platforms. Both architectures employ hardware and software prefetching: hardware prefetching is automatically enforced without explicit instructions, while software prefetching can be implemented using intrinsics like \textit{PLDL1KEEP} on Kunpeng 920 CPUs or \textit{\_mm\_prefetch} on x86 platforms. Memory optimizations like huge pages and NUMA are similar, where huge pages can help improve translation lookaside buffer (TLB) coverage for large datasets and NUMA-aware allocation can help improve memory locality.

\section{The KBest Library}\label{sec:algo}

 \begin{figure}[!t]
 	\centering
 	\includegraphics[width=\columnwidth]{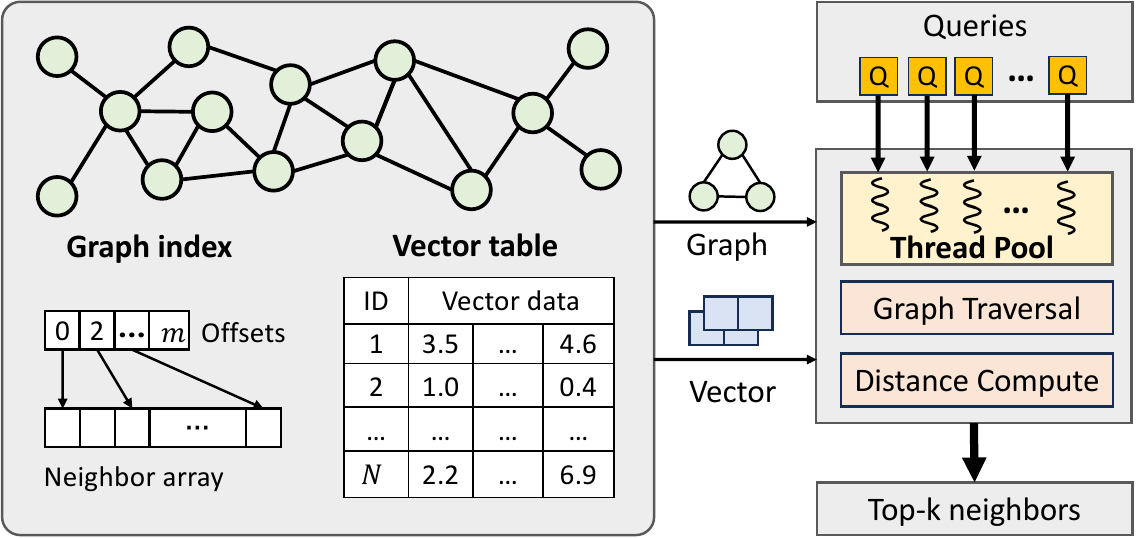}
 	\vspace{-5mm}
 	\caption{The memory layout and workflow of \name{}}
 	\label{fig:layout}
 	\vspace{-5mm}
 \end{figure}

\stitle{Overview} \name{} is an optimized graph-based ANNS algorithm specifically tuned for Kunpeng 920 CPUs, leveraging unique hardware features and algorithmic enhancements to achieve efficient search. As shown in Figure~\ref{fig:anns-system}, the system employs an in-memory architecture with two key components: (1) a CSR-formatted graph index with fixed out-degree $M$ (null-padded for uniform access), and (2) vector data stored in flattened, dimension-padded 1D arrays for memory alignment. \name{} utilizes a dynamic thread pool to automatically distribute incoming queries to idle worker threads, enabling concurrent processing while maintaining the Kunpeng architecture's full computational potential through both hardware-aware adaptations and general algorithmic optimizations.

  \begin{figure}[!t]
 	\centering
 	\includegraphics[width=0.55\columnwidth]{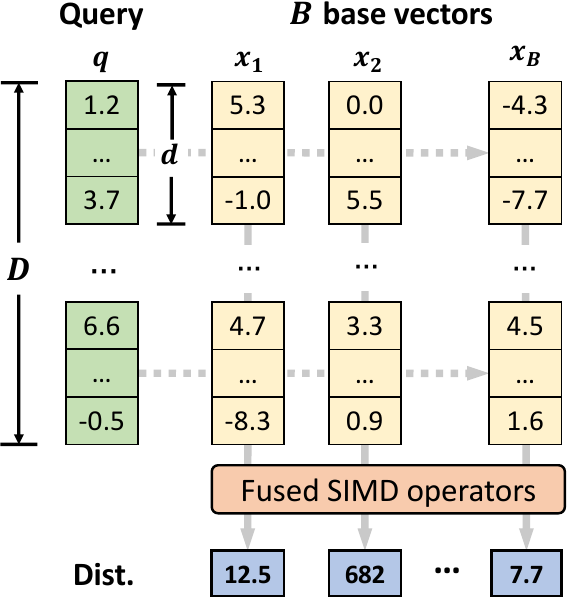}
 	\caption{The abstraction of SIMD accelerated operators of fused 1-to-$B$ distance computation}
 	\label{fig:simd-op}
    \vspace{-5mm}
 \end{figure}
\subsection{Kunpeng-aware Optimizations}\label{subsec:hardware-optimization}

\stitle{SIMD accelerated distance computation} Distance computation is the basic operation for graph-based ANNS, primarily executed when evaluating neighbors against query vectors to update candidate sets. While developers typically leverage architecture-specific SIMD extensions (namely AVX/AVX512 on x86 and NEON/SVE on ARM), \name{} introduces several Kunpeng hardware-specific optimizations. As showed in Figure~\ref{fig:simd-op}, we optimize SIMD instruction parallelism by exploiting the Kunpeng 920 CPU's multi-issue architecture, which enables concurrent execution of independent SIMD operations. In particular, we transform scalar 1-to-1 distance calculations into batched 1-to-$B$ vectorized operations, enabling up to 16 parallel distance computations per cycle when data dependencies permit. This approach fully saturates 128-bit NEON registers to maximize CPU utilization while amortizing memory access latency through query vector reuse across $B$ database items. For smaller workloads where batched processing is inefficient, we implement pipelined segmented accumulation for 1-to-1 distance calculations through carefully scheduled instruction streams. Second, we leverage some effective built-in fused operators to combine some fundamental operations into single SIMD intrinsics. For example, with float-type vectors, we utilize the \texttt{vmlaq\_f32} NEON instruction to fuse multiply-accumulate operations, reducing instruction count and improving pipeline efficiency.

\stitle{Data prefetch} Modern CPUs employ a hierarchical memory architecture where data must traverse multiple cache levels before reaching registers for computation: from main memory to L3 cache, then L2, and finally to L1. While modern processors implement hardware prefetch that predicatively load memory blocks based on access patterns, these mechanisms prove inadequate for graph-based ANNS algorithms. The irregular access patterns inherent to graph traversal due to the random access following the outgoing edges of the graph result in significant memory bottlenecks. To address this challenge, \name{} introduces a pipeline software prefetching strategy that loads data ahead of computation. 

\begin{figure}[!t]
 	\centering
 	\includegraphics[width=\columnwidth]{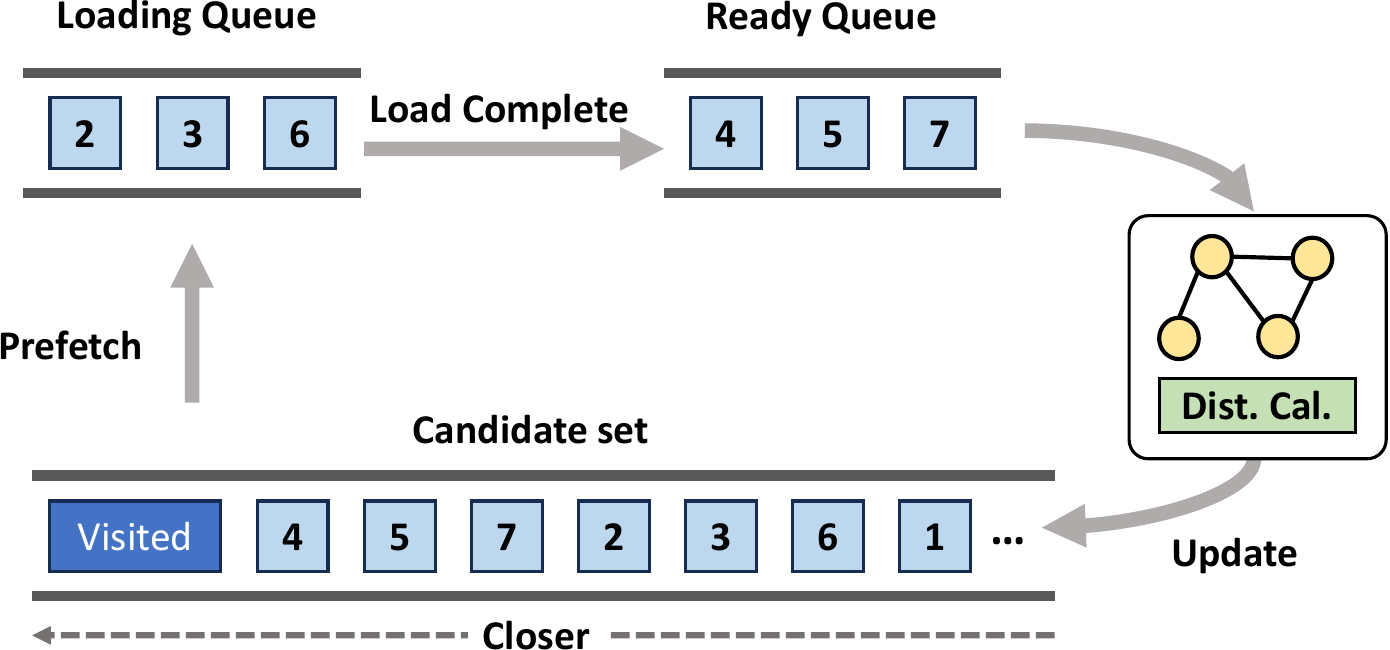}
 	\vspace{-5mm}
 	\caption{The workflow of \name{}'s prefetch strategy}
 	\label{fig:prefetching}
 	\vspace{-5mm}
 \end{figure}

Figure ~\ref{fig:prefetching} shows our prefetch strategy in the search process, where \name{} prefetches the adjacent list and vector data of the top priority node in the candidate set. This is because these nodes represent the immediate traversal targets in subsequent iterations. In each search iteration upon extracting the nearest node from the priority queue, the system prefetches a batch of $B$ neighbor nodes (where $B$ is determined by cache constraints) while concurrently processing the current node's neighbors. This batch prefetching continues until accumulating $B$ neighbors, at which point \name{} performs a batched one-to-many distance computation and inserts qualifying neighbors back into the candidate set. The prefetch batch size $B$ is determined by the cache-aware formulation:
\begin{equation}
B = \left\lfloor \frac{\alpha \cdot C_{\text{L1d}}}{d \cdot s} \right\rfloor
\label{eq:batch_size}
\end{equation}
where $C_{\text{L1d}}$ denotes the per-thread L1 data cache size, $d$ represents the vector dimensionality, $s$ is the element size in bytes (e.g., 4 bytes for float32), and $\alpha$ is the cache allocation ratio for prefetching (typically 0.5 for optimal compute-prefetch overlap).

Our implementation leverages ARM's low-level prefetch instruction through inline assembly assembly \textit{asm volatile("prfm PLD1KEEP, [\%0]" :: "r"(address))}. Here \textit{prfm} is an ARMv8 assembly instruction for cache prefetching, \textit{PLD1KEEP} operand specifies a long-term prefetch policy, instructing the memory subsystem to retain the prefetched data in cache hierarchy rather than treating it as transient and \textit{address} is the aligned memory location targeted for prefetching. This approach effectively bridges the latency gap between unpredictable memory accesses and computational pipelines.

\stitle{Memory management} \name{} implements dual-layer memory optimization to address the bandwidth-bound nature of graph-based ANNS on Kunpeng processors: At the virtual memory level, the system enforces 2MB huge page allocation through explicit 
\texttt{madvise(MADV\_HUGEPAGE)} directives and the system-level configure \texttt{/sys/kernel/mm/transparent\_hugepage/enabled}. This addresses the performance penalty caused by conventional 4KB page sizes, which induces significant Translation Lookaside Buffer (TLB) misses during random graph traversal, which is a critical bottleneck where adjacency lists and feature vectors may span hundreds of memory pages. Through contiguous physical mappings of huge pages, \name{} can reduce TLB miss rates while improving row buffer hit rates on Kunpeng's 8-channel DDR4 memory subsystem.

At the cache level, \name{} guarantees 64-byte aligned memory allocation for all critical data structures (including graph edges and feature vectors) via \textit{std::align\_alloc}, with each vector dimension padded to cache line boundaries. This alignment serves two purposes: (i) eliminating cross-cache-line access penalties during SVE/NEON vectorized distance computations, and (ii) minimizing cache coherence overhead through natural partition alignment.

\subsection{Algorithmic Optimizations}\label{subsec:algo-optimization}

\stitle{Index refinement} \name{} builds upon state-of-the-art graph construction strategies (e.g., NSG, SSG and Vamana) through a three-phase pipeline that enhances both construction efficiency and search performance. First, a k-nearest neighbors (kNN) graph is constructed where each vertex's neighborhood contains its nearest vectors from the dataset. We employ the advanced RNNDescent~\cite{rnndescent} algorithm to generate the initial kNN graph for its high efficiency and quality guarantee. Second, we refine each vertex's neighborhood based on search results from the initial kNN graph, ensuring the edges cover diverse directions in the vector space. To accommodate different dataset characteristics, we support several edge selection strategies in SOTA graph index. For instance, HNSW, NSG with their distance-based strategies and SSG with its angle-based selection rules. Additionally, we introduce a novel iterative refinement strategy, during each iteration, we expand candidate to include all 2-hop neighbors for each node and reapply our edge selection rules. This process continues for $F$ iterations until the graph stabilizes or the construction time budget is exhausted. This optimization can help shorten the search path on graph and improve efficiency.

\stitle{Graph reordering} Graph reordering is a cache optimization technique that improves memory locality by placing neighboring nodes in consecutive or near-consecutive memory locations. When a node and its associated vector data are loaded into memory, modern CPUs often prefetch adjacent memory blocks. By reordering nodes so that likely traversal paths correspond to spatially adjacent memory, the algorithm benefits from higher cache hit rates.

To formally define the optimization goal of graph reordering, we adopt the well-known graph bandwidth minimization problem. Given a graph \( G = (V, E) \) with vertices \( V = \{v_1, v_2, \ldots, v_n\} \), the objective is to find a bijection \( \pi: V \rightarrow \{1, 2, \ldots, n\} \) that maps each node to a unique memory position such that the maximum distance between connected nodes is minimized:

\begin{equation}
\min_{\pi} \max_{(v_i, v_j) \in E} |\pi(v_i) - \pi(v_j)|
\end{equation}

Here, \( \pi(v_i) \) and \( \pi(v_j) \) denote the memory positions of nodes \( v_i \) and \( v_j \). Minimizing this objective places connected nodes closer in memory, improving spatial locality and reducing cache misses during graph traversal.

Unfortunately, the graph bandwidth minimization problem is known to be NP-hard, making it unlikely to admit a polynomial-time exact solution. While heuristic methods such as Cuthill-McKee~\cite{graphreordercm} and Gorder ~\cite{gorder} have been proposed, we observe that these remain suboptimal for graph index traversal in ANNS. This limitation stems from fundamental incompatibilities: ANNS graphs exhibit small-world properties characterized by densely clustered local connections and sparse long-range shortcuts, forming heterogeneous topologies that conventional algorithms fail to preserve. Specifically, Cuthill-McKee is designed for matrix bandwidth minimization—disrupts critical ANNS shortcuts during BFS traversal through long-range label jumps, artificially elongating search paths. Similarly, Gorder's cache-locality optimization may forcibly co-locate topologically connected but geometrically distant nodes in high-dimensional space, violating the underlying data geometry essential for ANNS performance.

Considering this important long-range shortcut edges, we propose a specialized graph reordering algorithm tailored for ANNS in Algorithm~\ref{alg:graph-reorder}. In algorithm~\ref{alg:graph-reorder} we first constructs a minimum spanning tree (MST) $T$ from the original graph $G$ to establish connectivity and select the entry node $e$. Second, line 5-16 computes subtree sizes (met(v)) for all nodes via an iterative depth-first search (DFS), where each node's metric aggregates its subtree cardinality. Finally, line 17-23 performs a prioritized traversal that processes nodes in descending order of their subtree sizes, effectively clustering densely connected regions together in memory. This approach reduces cache misses by ensuring frequently co-accessed nodes (those in large subtrees) are stored contiguously.

\begin{algorithm}[t]
\caption{Graph Node Reordering for Memory Locality}
\label{alg:node_reordering}
\begin{algorithmic}[1]
\State \textbf{Input:} Original graph $G$, entry node $e$
\State \textbf{Output:} Reordered node sequence $S$

\State Generate minimum spanning tree (MST) $T$ from $G$ 
\State Select the entry node $e$ as the root of $T$

\State Initialize $\text{met}(v) \gets 1, \forall v \in T$ 
\State Initialize empty stack and visited set
\State Push $(e, \text{False})$ onto stack 

\While{stack is not empty}
    \State $(u, processed) \gets \text{stack.pop}()$
    \If{processed}
        \For{each child $v$ of $u$ in $T$}
            \State $\text{met}(u) \gets \text{met}(u) + \text{met}(v)$
        \EndFor
    \Else
        \State Push $(u, \text{True})$ onto stack
        \For{each child $v$ of $u$ in $T$ (in reverse order)}
            \State Push $(v, \text{False})$ onto stack \Comment{DFS visit order}
        \EndFor
    \EndIf
\EndWhile

\State Initialize empty priority queue $Q$ and list $S$
\State Push root node into $Q$
\While{$Q$ is not empty}
    \State $u \gets Q.\text{pop}()$ \Comment{Dequeue max $\text{met}(v)$}
    \State $S.\text{append}(u)$
    \For{each child $v$ of $u$ in $T$}
        \State $Q.\text{push}(v, \text{key}=\text{met}(v))$
    \EndFor
\EndWhile
\State \Return $S$
\end{algorithmic}
\label{alg:graph-reorder}
\end{algorithm}

\stitle{Early termination} Graph-based ANNS algorithms utilize Best-First Search (BFS) to traverse the graph structure while maintaining a fixed-size candidate list $L$ to track potential nearest neighbors. The search terminates when all nodes in $L$ have been visited, with the parameter L critically determining both search scope and recall accuracy. We find that the candidate list exhibits low utilization rates that may cause  numerous invalid search paths, resulting in substantial computational overhead. To address this inefficiency, we propose an early termination algorithm that dynamically determines when to halt the search. The algorithm tracks two critical metrics: The insertion position $p \in \mathbb{Z}^+$ of each new candidate in the candidate set and the number of consecutive insertions occurring beyond a threshold position $t$. The search terminates when consecutive insertions beyond $t$ exceed $\tau_{\text{max}}$:

\begin{equation}
\text{EarlyTerm.}(t, \tau_{\text{max}}) := \left[\sum_{i=k-\tau_{\text{max}}}^{k} \mathbb{I}(p_i > t)\right] \geq \tau_{\text{max}}
\label{eq:earlystop}
\end{equation}

The early-termination heuristic relies on the observation that when consecutive unvisited nodes consistently rank near the end of the candidate list (i.e., farthest from the query), the search is likely diverging from the query's neighborhood. Here the threshold position $t$ and patience $\tau_{\text{max}}$ are tunable parameters and their optimal values depend heavily on the dataset characteristics and recall requirements. To determine the best configuration, we perform a two-stage search with dry-run queries. In practice, we initialize with $t$ as about 60\% of $L$, the total search list size in BFS, and then conduct binary search for $\tau_{\text{max}}$ under the given recall constraint. To further optimize search performance while maintaining recall, we explore reducing $t$ from 60\% down to 30\% of $L$, identifying the setting that maximizes search speed without compromising recall.

\stitle{Vector quantization} This technique compresses high-dimensional data by mapping vectors to discrete codewords from a learned codebook. The process dramatically reduces storage and computation costs while preserving relative similarity, making it ideal for large-scale search systems that operate on compressed representations instead of raw data. Two widely-used variants include Product Quantization (PQ) and Scalar Quantization (SQ), which PQ divides the vector space into orthogonal subspaces for independent quantization, and SQ operates by independently quantizing each vector component to a fixed scalar range.

Notably, \name{}'s quantization module is implemented as a standalone component with well-defined interfaces, enabling seamless integration of vector quantization algorithms to further enhance retrieval efficiency and accuracy without architectural modifications. This modular design ensures forward compatibility with emerging quantization techniques while maintaining the system's core optimization pipeline.

\section{API and Use Cases of \name{}}\label{sec:imple}
In this section, we present \name{}'s API design and demonstrate its deployment options, including usage as a standalone library and integration with Milvus~\cite{milvus} as an ANNS algorithm component.

\subsection{The API of \name{}}

\begin{table}[!t]
\caption{Key APIs of \name{}}
\label{tab:api}
\begin{tabular}{@{}ccp{3cm}}
\toprule
\textbf{Step} & \textbf{\name{} API} & \multicolumn{1}{c}{\textbf{Interpretation}} \\ \midrule
\multirow{2}{*}{\textbf{\begin{tabular}[c]{@{}c@{}}Parameter \\ prepare\end{tabular}}} & \multirow{2}{*}{\textit{KBest(config)}} & \multirow{2}{\linewidth}{Initalizea \name{} with detailed configuration} \\
 &  &  \\ \midrule
\multirow{2}{*}{\textbf{\begin{tabular}[c]{@{}c@{}}Index\\ construct\end{tabular}}} & \multirow{2}{*}{\textit{Add(n, x)}} & \multirow{2}{\linewidth}{build the graph index with n input index} \\
 &  &  \\ \midrule
\multirow{2}{*}{\textbf{\begin{tabular}[c]{@{}c@{}}Query\\ process\end{tabular}}} & \multirow{2}{*}{\textit{Search(nq, q, k, nt)}} & \multirow{2}{\linewidth}{search top-k NN of $nq$ queries with $nt$ threads} \\
 &  &  \\ \bottomrule
\end{tabular}
\vspace{-3mm}
\end{table}

\name{} is implemented in C++ with approximately 9K lines of code, and provides user-friendly interfaces in both C++ and Python. Like other mainstream libraries, it organizes the ANNS workflow into three stages: parameter preparation, index construction, and query processing. In the first stage, users initialize a \name{} instance with specified parameters. During index construction, base vectors and graph-building parameters are provided to build the index. In the final stage, \name{} answers queries using the constructed graph and vector data. Table~\ref{tab:api} summarizes the key APIs supporting these steps. To avoid redundant index building, \name{} also supports saving and loading pre-built graph indices.

\name{} seamlessly integrates with vector databases like Milvus~\cite{milvus} through the Knowhere computation layer, which orchestrates the entire workflow by calling \name{}'s specific interfaces. During the build phase, Knowhere invokes \name{}'s graph construction API to incrementally build the index, which is then serialized along with the user data for storage. When executing queries, Knowhere manages the complete search pipeline by loading indices through \name{}'s deserialization interface, performing efficient top-k search via its optimized query interface, and delivering results to Milvus' distributed query coordinator. Through this tight integration, Knowhere ensures \name{} maintains its native performance while fully leveraging Milvus' distributed architecture and hardware acceleration capabilities.

\subsection{Use Cases of \name{}}

\name{} has been widely adopted in large-scale industrial applications, demonstrating exceptional scalability and efficiency in real-world scenarios. Notably, it has been successfully deployed in top-tier social media, e-commerce, and food delivery platforms, processing tens of millions of daily queries across massive server clusters while maintaining millisecond-level latency. In social media applications, user-generated content including images and short videos is automatically encoded into dense vector representations through advanced AI models, while user preferences are simultaneously converted into query vectors, enabling real-time personalized content recommendations through high-performance similarity search.

For e-commerce platforms, the system efficiently transforms product information including titles, descriptions and images into item vectors, while converting user search queries (such as "wireless headphones") into corresponding query vectors. This allows \name{} to instantly retrieve the most relevant products from its large-scale indexes, delivering low-latency recommendations even during peak traffic periods, handling millions of concurrent requests. In the food delivery sector, restaurant menus, dish images and location-based user queries are intelligently mapped to a unified vector space. \name{}'s optimized architecture enables instantaneous search and matching of dishes with consistent response times less than 5ms.

Additionally, \name{} has been integrated into mainstream vector databases including Milvus~\cite{milvus} and OpenGauss~\cite{opengauss}, enhancing their vector search capabilities. Beyond commercial platforms, \name{} serves national infrastructure projects by enabling rapid similarity search across satellite imagery and sensor data vectors, while telecom companies leverage it to match network patterns and troubleshoot issues from encoded operational data. These mission-critical deployments utilize thousands of servers processing tens of millions of queries daily while maintaining high availability, security, and performance standards.

\section{Experimental Evaluation}\label{sec:exp}

\begin{table}[!t]
\centering
	\caption{Statistics of the vector datasets used for experiments.}
    \label{tab:dataset}
        \begin{tabular}{cccc}
        \hline
        \textbf{Dataset}  & \textbf{Num} & \textbf{Dim} & \textbf{Similarity} \\ \hline
        Glove            & 1M           & 100          & Angular             \\ 
        Deep             & 10M          & 96           & Angular             \\ 
        Text-to-Image       & 10M          & 200          & Inner-Product       \\ 
        BigANN           & 100M         & 128          & L2                  \\ \hline
        \end{tabular}
    \vspace{-3mm}
\end{table}

\begin{figure*}[!t]
	\centering
	\includegraphics[width=1.9\columnwidth]{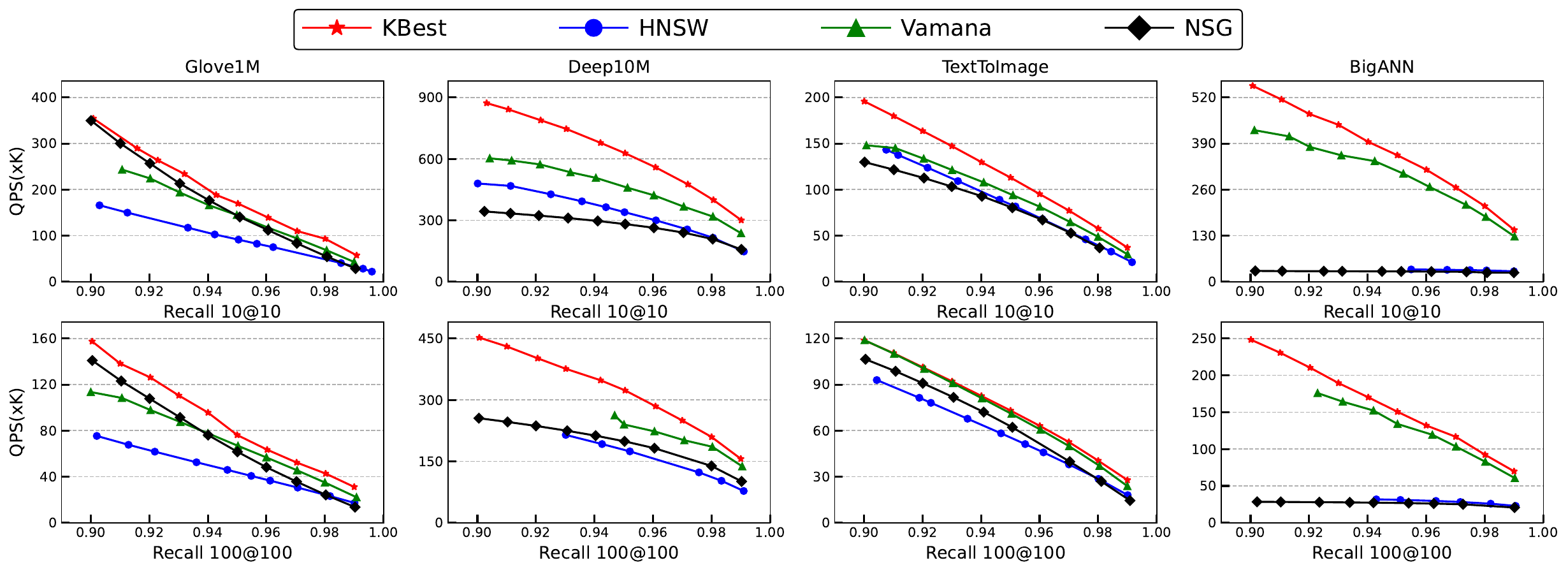}
	\vspace{-3mm}
	\caption{QPS vs. Recall of \name{} on Kunpeng 920 compared with baselines on AMD 9654}
	\label{fig:main-exp}
\end{figure*}

    \begin{table*}[!t]
    \caption{The QPS of \name{} on Kunpeng 920 and baselines on AMD 9654 at recall 0.95}
    \label{tab:result-recall95}
    \begin{tabular}{c|cccc|cccc} 
    \toprule[1.5pt]
    \multicolumn{1}{c|}{\textbf{}} & \multicolumn{4}{c|}{\textbf{QPS (xK) Recall@10=0.95}} & \multicolumn{4}{c}{\textbf{QPS (xK) Recall@100=0.95}} \\ 
    \midrule 
    \textbf{Datasets}             & \textbf{Glove}     & \textbf{Deep}     & \textbf{T2Img}     & \textbf{B-ANN}     & \textbf{Glove}     & \textbf{Deep}     & \textbf{T2Img}     & \textbf{B-ANN}     \\ 
    \midrule
    \textit{NSG}                  & 140      & 282      & 80           & 29      & 62       & 199      & 62           & 27      \\ 
    \textit{HNSW}                 & 91       & 340     & 82           & 35      & 41       & 174      & 51          & 31      \\ 
    \textit{Vamana}               & \underline{145}      & \underline{459}     & \underline{94}           & \underline{305}     & \underline{67}       & \underline{240}      & \underline{71}           & \underline{134}     \\ 
    \textit{KBest}                & \textbf{170}      & \textbf{628}     & \textbf{113}          & \textbf{357}     & \textbf{76}       & \textbf{323}      & \textbf{73}           & \textbf{151}     \\ \midrule
    Comparison           & 1.17x        & 1.37x        & 1.20x            & 1.17x       & 1.14x        & 1.35x        & 1.03x            & 1.12x       \\ 
    \bottomrule[1.5pt]
    \end{tabular}
    \end{table*}

We conduct comprehensive experiments to evaluate \name{}'s performance on Huawei Kunpeng 920 CPUs against state-of-the-art (SOTA) graph-based ANNS algorithms on x86 platforms. We further conduct an ablation study that quantifies the individual contribution of each design to the performance gains.

\subsection{Experiment Settings}\label{subsec:exp setting}

\stitle{Datasets} We evaluate \name{} on four standard benchmark datasets: Glove~\cite{glove}, Deep~\cite{deep}, Text-to-Image~\cite{yandextexttoimage}, and BigANN~\cite{BigANN2021,sift}. We summarize the dataset characteristic in Table~\ref{tab:dataset}. To comprehensively show the performance of varying scales and cases, we use a sampled 1M subset of dataset of Glove and a 10M subset of dataset of Deep. In Text-to-Image dataset, the distribution of queries is different from the input data and we verify the capability of \name{} of handling out-of-distribution dataset. Finally, to show the scalability of \name{}, we evaluate the performance on the BIGANN of 100M scale.

\stitle{Baselines} We compare our \name{} approach with three representative graph index types: HNSW, NSG, and Vamana, using their respective state-of-the-art implementations which includes:
\squishlist
    \item \textbf{HNSW}~\cite{faiss}: We use the Faiss library for HNSW. Faiss is developed by Meta's Fundamental AI Research (FAIR) team and provides highly optimized implementations for most useful SOTA vector search algorithms. The implementation leverages multiple acceleration libraries including MKL and OpenBLAS, combined with low-level optimizations of graph structures and search procedures to achieve outstanding query performance. 
    \item \textbf{NSG}~\cite{PyGlass}: For NSG implementation, we employ Zilliz's Pyglass library, a lightweight solution that implements NSG and other graph indexes without any third-party dependencies. The implementation demonstrates sophisticated memory management and optimized data structure design that significantly reduces memory footprint while maintaining high search accuracy. 
    \item \textbf{Vamana}~\cite{diskann}: We utilize the in-memory version from Microsoft's DiskANN project, which pioneered an innovative two-pass edge selection strategy with adjustable parameter $\alpha$ to optimize graph density. DiskANN's in-memory implementation preserves all Vamana's algorithmic advantages while optimizing for RAM-based operation through compressed data representation and efficient memory access patterns, achieving excellent query performance in memory-resident environments.
\squishend

For each method, we adopt the default graph construction parameters recommended by their authors, including the number of neighbors per node $M$ and the search list size $L$ during index building. After constructing the graph indices in advance for each dataset, we vary the search list size $efs$ to explore different throughput–recall trade-offs.

\stitle{Platform and performance metrics} We evaluate \name{} on an ARM-based platform and compare it against other baselines running on x86 platforms. The primary ARM testbed is equipped with 2.9GHz Huawei Kunpeng 920 processors, running OpenEuler 22.04. For the SOTA baselines, we use a powerful x86 platform with an AMD EPYC 9654 processor overclocked to 3.7GHz, running Ubuntu 22.04. To ensure a fair comparison, we enable hyper-threading and utilize all available threads to handle queries in parallel.

We focus on evaluating both the efficiency and accuracy of each ANNS algorithm. Specifically, we use full-machine query-per-second (QPS) as the performance metric, and recall@10 and recall@100 (i.e., 10@10 and 100@100 recall) as the accuracy metrics.

\subsection{Main Results}\label{subsec:exp main}

Figure~\ref{fig:main-exp} shows the QPS-recall curves comparing \name{} with other x86-based baselines, while Table~\ref{tab:result-recall95} provides detailed QPS measurements under a 0.95 recall requirement. The results demonstrate that in the high-recall range of 0.90-1.0, \name{} achieves 1.04x–1.34x higher performance than the best baseline (Vamana) and up to 12.6x improvement over other baselines across all four datasets. This performance advantage stems from Kunpeng-specific hardware optimizations combined with algorithmic enhancements to graph index quality and search efficiency.

We first examine the a million-scale Glove dataset. As showed in Figure~\ref{fig:main-exp}, under low recall requirements, \name{} and other baselines retrieve correct results efficiently, resulting in only modest performance gains: \name{} achieves 1.17× and 1.14× higher QPS than the best baseline Vamana at recall@10=0.95 and recall@100=0.95, respectively in Table~\ref{tab:result-recall95}. However, as the recall target increases toward near-exact, the QPS of all methods drops significantly. In this high-recall regime, \name{}'s advantage becomes more pronounced, reaching up to 1.4× the performance of Vamana. This is attributed to the dense distribution of the Glove dataset that achieving near-exact recall requires searching a large number of vectors around the query, which introduces many redundant paths. \name{}’s early termination mechanism effectively prunes these redundant paths, leading to improved efficiency.

Next, we evaluate the significantly larger Deep dataset. Despite its size, the overall QPS is notably higher. This is because Deep better reflects the distribution of real-world datasets, leading to shorter search paths in the graph index across all methods. Additionally, the dataset has relatively low dimensionality, which amplifies the impact of memory access bottlenecks commonly seen in graph-based ANNS. To address this, \name{} incorporates several memory efficiency optimizations, including prefetch pipelines and an optimized memory layout to reduce cache misses, as well as graph reordering techniques that convert random access patterns into relatively sequential ones. These strategies significantly improve memory locality and utilization. As a result, \name{} achieves up to 1.45× times QPS compared to the best baseline Vamana.

To evaluate the robustness of \name{}, we assess its performance on an out-of-distribution (OOD) scenario using the Text-to-Image dataset, where the query vector distribution differs from that of the training data used to construct the graph index. As shown in Table~\ref{tab:result-recall95}, \name{} maintains superior performance at recall@10=0.95 and recall@100=0.95, achieving 1.20× and 1.03× QPS respectively compared to the best baseline Vamana. Figure~\ref{fig:main-exp} demonstrates that across the high-recall range of 0.9-1.0, \name{} delivers consistently better performance, with average QPS of 1.22× for recall@10 and 1.05× for recall@100 over Vamana, indicating its strong generalization capability under distribution shift.

To evaluate the scalability of \name{} and other baselines, we conduct experiments on the 100M-scale BigANN dataset. Compared to previous small datasets, BigANN presents more severe challenges for graph traversal due to its larger size and the increased bottleneck of random memory access. As shown in Figure~\ref{fig:main-exp} and Table~\ref{tab:result-recall95}, baselines HNSW by Faiss and NSG by PyGlass exhibit significantly degraded performance, with only marginal improvements in recall as the search parameter $efs$ increases. This limitation arises because these methods are primarily optimized for smaller datasets. At the billion-scale, their performance is heavily constrained by memory access overhead, dominating the total search time. In contrast, Vamana by DiskANN is optimized for disk-based large-scale search, achieving much higher QPS on BigANN. Notably, \name{} still outperforms Vamana, delivering average QPS of 1.22× and 1.24× at recall@10 and recall@100.

\subsection{Ablation Study}\label{subsec:exp minor}

\begin{figure}[!t]
	\centering
	\includegraphics[width=\columnwidth]{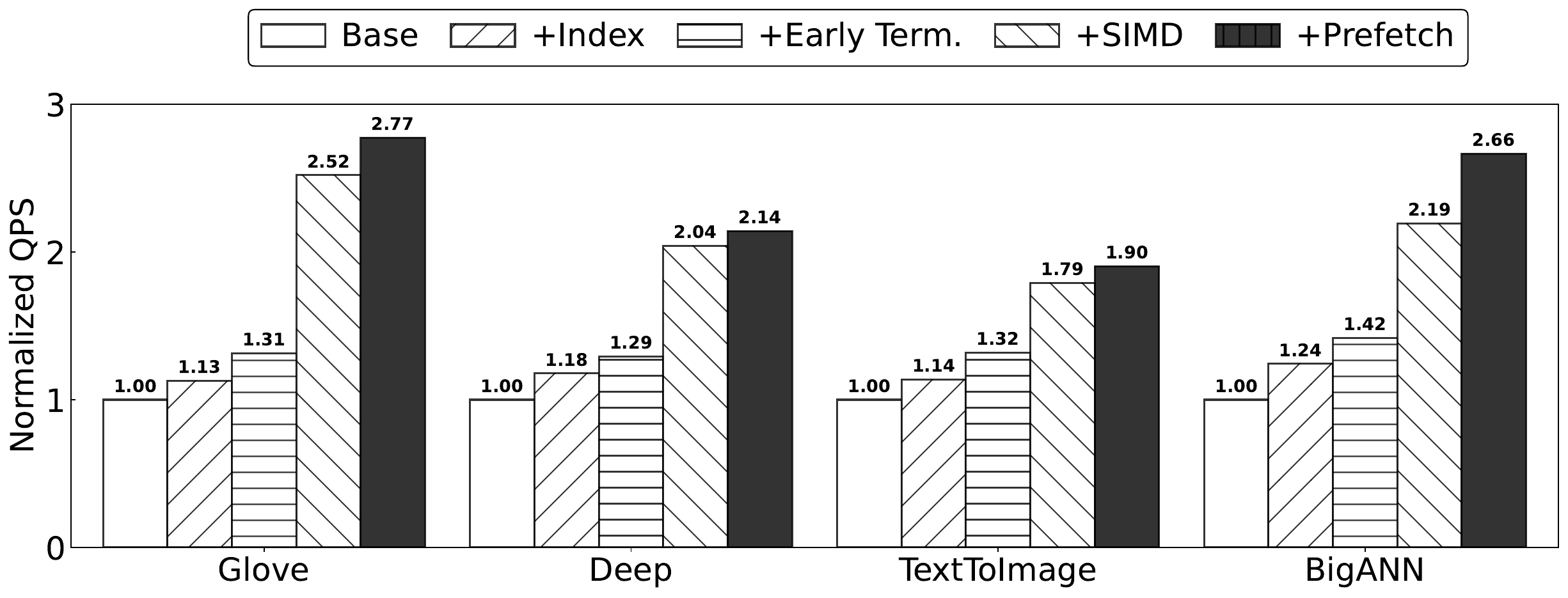}
	\vspace{-6mm}
	\caption{Ablation study of \name{}}
	\label{fig:micro-exp}
	\vspace{-5mm}
\end{figure}

To assess the impact of individual optimizations, we conduct an ablation study by progressively enhancing the base version of \name{}: starting with the unoptimized baseline(Base), we first incorporate graph index refinements (+Index), then introduce early termination (+Early Term.), followed by SIMD acceleration (+SIMD), and finally integrate prefetching (+Prefetch) to arrive at the fully optimized \name{}.

As shown in Figure~\ref{fig:micro-exp}, all four optimizations demonstrate consistent performance gains across these 4 datasets. The graph index optimization yields improvements ranging from 12.8\% (Glove) to 24.3\% (BigANN), with greater benefits observed at larger scales due to increased optimization potential in the index structure. Early termination achieves performance gains of up to 16\% on datasets like Glove and Text-To-Image. This improvement stems from its ability to prune redundant search paths, particularly effective for non-standard data distributions (e.g., Glove) and out-of-distribution cases (e.g., Text-to-Image). The most significant improvements come from SIMD optimization, delivering average 60\% and up to 92\% acceleration by addressing the computational bottleneck in distance calculations through our proposed fused SIMD operators and 1-to-$B$ vectorized operations, which maximize parallelization efficiency on Kunpeng CPUs. Finally, prefetching optimization specifically targets memory access bottlenecks in large graphs, achieving more than 20\% performance boost on the billion-scale BigANN dataset.

\section{Related Work}\label{sec:related-work}

\stitle{Vector indexes}  Over the past few decades, various vector indexing methods have been developed to efficiently solve the Approximate Nearest Neighbor Search (ANNS) problem: the hashing-based methods (e.g., Local Sensitive Hashing (LSH)~\cite{lsh} and Spectral Hashing~\cite{spectral_hashing}) split the dataset and index vectors via hash tables using distance-preserving hash functions. Tree-based methods (e.g., KD-tree~\cite{kdtree} and R-tree~\cite{rtree}) recursively organize vectors into hierarchical tree structures with spatial partitioning. Inverted File (IVF) methods, including IVFPQ~\cite{ivfpq} and Scann~\cite{scann}, first partition the dataset via clustering and build inverted indices for coarse-to-fine search. Our \name{} algorithm is a graph-based ANNS method that constructs optimized proximity graphs for greedy traversal. Through custom graph refinement techniques, it achieves superior search efficiency and accuracy compared to state-of-the-art graph indexes including HNSW~\cite{hnsw}, Vamana~\cite{diskann}, and SSG~\cite{ssg}.

\stitle{Vector search libraries} The exponential growth of vector data has driven significant advances in modern search systems. These libraries combine innovative indexing structures with hardware-conscious optimizations to deliver high performance under rigorous recall requirements. FAISS~\cite{faiss}, developed by Meta, implements IVF, PQ, and HNSW indices along with clustering algorithms for efficient vector search, while leveraging GPU acceleration for enhanced throughput. Microsoft's DiskANN~\cite{diskann} employs SSD-optimized Vamana graphs with PQ compression, reducing memory overhead through intelligent disk-memory hierarchy management. Google's Scalable Nearest Neighbors (ScaNN)~\cite{scann} utilizes anisotropic vector quantization~\cite{avq} and SOAR~\cite{soar} techniques to optimize both inner-product and distance-based searches without compromising recall accuracy. While these state-of-the-art systems are primarily optimized for x86 architectures, our \name{} is specifically designed for Kunpeng 920 CPUs, incorporating both hardware-aware and algorithmic optimization to outperform other libraries on x86 CPUs.

\section{Conclusions}\label{sec:conclusion}
We present \name{}, an efficient graph-based vector search library specifically optimized for Huawei Kunpeng 920 CPUs. We find that existing state-of-the-art vector search libraries primarily target x86 platforms and exhibit suboptimal performance on ARM architectures. To bridge this performance gap, we implement ARM-specific hardware optimizations, including accelerated SIMD operators, software prefetching, and memory-related enhancements. Additionally, we introduce general algorithmic improvements such as graph index refinement and early termination during the search process. \name{} offers a user-friendly API and has been widely adopted, both within our internal business and by external enterprises.

\bibliographystyle{ACM-Reference-Format}
\bibliography{ref}


\begin{thebibliography}{49}


\ifx \showCODEN    \undefined \def \showCODEN     #1{\unskip}     \fi
\ifx \showISBNx    \undefined \def \showISBNx     #1{\unskip}     \fi
\ifx \showISBNxiii \undefined \def \showISBNxiii  #1{\unskip}     \fi
\ifx \showISSN     \undefined \def \showISSN      #1{\unskip}     \fi
\ifx \showLCCN     \undefined \def \showLCCN      #1{\unskip}     \fi
\ifx \shownote     \undefined \def \shownote      #1{#1}          \fi
\ifx \showarticletitle \undefined \def \showarticletitle #1{#1}   \fi
\ifx \showURL      \undefined \def \showURL       {\relax}        \fi
\providecommand\bibfield[2]{#2}
\providecommand\bibinfo[2]{#2}
\providecommand\natexlab[1]{#1}
\providecommand\showeprint[2][]{arXiv:#2}

\bibitem[ann-benchmarks Team(2021)]%
        {BigANN2021}
\bibfield{author}{\bibinfo{person}{Big ann-benchmarks Team}.} \bibinfo{year}{2021}\natexlab{}.
\newblock \bibinfo{title}{BigANN Benchmark: NeurIPS'21 Track}.
\newblock \bibinfo{howpublished}{\url{https://big-ann-benchmarks.com/neurips21.html}}.
\newblock
\newblock
\shownote{Accessed: 2023-11-15}.


\bibitem[Babenko and Lempitsky(2016)]%
        {deep}
\bibfield{author}{\bibinfo{person}{Artem Babenko} {and} \bibinfo{person}{Victor Lempitsky}.} \bibinfo{year}{2016}\natexlab{}.
\newblock \showarticletitle{Efficient indexing of billion-scale datasets of deep descriptors}. In \bibinfo{booktitle}{\emph{Proceedings of the IEEE Conference on Computer Vision and Pattern Recognition}}. \bibinfo{pages}{2055--2063}.
\newblock


\bibitem[Beckmann et~al\mbox{.}(1990)]%
        {rtree}
\bibfield{author}{\bibinfo{person}{Norbert Beckmann}, \bibinfo{person}{Hans{-}Peter Kriegel}, \bibinfo{person}{Ralf Schneider}, {and} \bibinfo{person}{Bernhard Seeger}.} \bibinfo{year}{1990}\natexlab{}.
\newblock \showarticletitle{The R*-Tree: An Efficient and Robust Access Method for Points and Rectangles}. In \bibinfo{booktitle}{\emph{Proceedings of the 1990 {ACM} {SIGMOD} International Conference on Management of Data, Atlantic City, NJ, USA, May 23-25, 1990}}, \bibfield{editor}{\bibinfo{person}{Hector Garcia{-}Molina} {and} \bibinfo{person}{H.~V. Jagadish}} (Eds.). \bibinfo{publisher}{{ACM} Press}, \bibinfo{pages}{322--331}.
\newblock
\href{https://doi.org/10.1145/93597.98741}{doi:\nolinkurl{10.1145/93597.98741}}


\bibitem[Chen et~al\mbox{.}(2024)]%
        {bge}
\bibfield{author}{\bibinfo{person}{Jianlv Chen}, \bibinfo{person}{Shitao Xiao}, \bibinfo{person}{Peitian Zhang}, \bibinfo{person}{Kun Luo}, \bibinfo{person}{Defu Lian}, {and} \bibinfo{person}{Zheng Liu}.} \bibinfo{year}{2024}\natexlab{}.
\newblock \showarticletitle{{BGE} M3-Embedding: Multi-Lingual, Multi-Functionality, Multi-Granularity Text Embeddings Through Self-Knowledge Distillation}.
\newblock \bibinfo{journal}{\emph{CoRR}}  \bibinfo{volume}{abs/2402.03216} (\bibinfo{year}{2024}).
\newblock
\showeprint[arXiv]{2402.03216}
\href{https://doi.org/10.48550/ARXIV.2402.03216}{doi:\nolinkurl{10.48550/ARXIV.2402.03216}}


\bibitem[Chen et~al\mbox{.}(2021)]%
        {spann}
\bibfield{author}{\bibinfo{person}{Qi Chen}, \bibinfo{person}{Bing Zhao}, \bibinfo{person}{Haidong Wang}, \bibinfo{person}{Mingqin Li}, \bibinfo{person}{Chuanjie Liu}, \bibinfo{person}{Zengzhong Li}, \bibinfo{person}{Mao Yang}, {and} \bibinfo{person}{Jingdong Wang}.} \bibinfo{year}{2021}\natexlab{}.
\newblock \showarticletitle{SPANN: Highly-efficient Billion-scale Approximate Nearest Neighborhood Search}. In \bibinfo{booktitle}{\emph{Advances in Neural Information Processing Systems}}, \bibfield{editor}{\bibinfo{person}{M.~Ranzato}, \bibinfo{person}{A.~Beygelzimer}, \bibinfo{person}{Y.~Dauphin}, \bibinfo{person}{P.S. Liang}, {and} \bibinfo{person}{J.~Wortman Vaughan}} (Eds.), Vol.~\bibinfo{volume}{34}. \bibinfo{publisher}{Curran Associates, Inc.}, \bibinfo{pages}{5199--5212}.
\newblock
\urldef\tempurl%
\url{https://proceedings.neurips.cc/paper_files/paper/2021/file/299dc35e747eb77177d9cea10a802da2-Paper.pdf}
\showURL{%
\tempurl}


\bibitem[Chen and Zou(2024)]%
        {genept}
\bibfield{author}{\bibinfo{person}{Yiqun Chen} {and} \bibinfo{person}{James Zou}.} \bibinfo{year}{2024}\natexlab{}.
\newblock \showarticletitle{GenePT: a simple but effective foundation model for genes and cells built from ChatGPT}.
\newblock \bibinfo{journal}{\emph{bioRxiv}} (\bibinfo{year}{2024}), \bibinfo{pages}{2023--10}.
\newblock


\bibitem[Cuevas{-}Tello et~al\mbox{.}(2013)]%
        {medicine3}
\bibfield{author}{\bibinfo{person}{Juan~Carlos Cuevas{-}Tello}, \bibinfo{person}{Daniel Hern{\'{a}}ndez{-}Ram{\'{\i}}rez}, {and} \bibinfo{person}{Christian~Alberto Garcia{-}Sepulveda}.} \bibinfo{year}{2013}\natexlab{}.
\newblock \showarticletitle{Support vector machine algorithms in the search of {KIR} gene associations with disease}.
\newblock \bibinfo{journal}{\emph{Comput. Biol. Medicine}} \bibinfo{volume}{43}, \bibinfo{number}{12} (\bibinfo{year}{2013}), \bibinfo{pages}{2053--2062}.
\newblock
\href{https://doi.org/10.1016/J.COMPBIOMED.2013.09.027}{doi:\nolinkurl{10.1016/J.COMPBIOMED.2013.09.027}}


\bibitem[Cuthill and McKee(1969)]%
        {graphreordercm}
\bibfield{author}{\bibinfo{person}{Elizabeth~H. Cuthill} {and} \bibinfo{person}{J. McKee}.} \bibinfo{year}{1969}\natexlab{}.
\newblock \showarticletitle{Reducing the bandwidth of sparse symmetric matrices}. In \bibinfo{booktitle}{\emph{Proceedings of the 24th national conference, {ACM} 1969, USA, 1969}}, \bibfield{editor}{\bibinfo{person}{Solomon~L. Pollack}, \bibinfo{person}{Thomas~R. Dines}, \bibinfo{person}{Ward~C. Sangren}, \bibinfo{person}{Norman~R. Nielsen}, \bibinfo{person}{William~G. Gerkin}, \bibinfo{person}{Alfred~E. Corduan}, \bibinfo{person}{Len Nowak}, \bibinfo{person}{James~L. Mueller}, \bibinfo{person}{Joseph~Horner III}, \bibinfo{person}{Pasteur S.~T. Yuen}, \bibinfo{person}{Jeffery Stein}, {and} \bibinfo{person}{Margaret~M. Mueller}} (Eds.). \bibinfo{publisher}{{ACM}}, \bibinfo{pages}{157--172}.
\newblock
\href{https://doi.org/10.1145/800195.805928}{doi:\nolinkurl{10.1145/800195.805928}}


\bibitem[Dmitry~Baranchuk(2021)]%
        {yandextexttoimage}
\bibfield{author}{\bibinfo{person}{Artem~Babenko Dmitry~Baranchuk}.} \bibinfo{year}{2021}\natexlab{}.
\newblock \bibinfo{title}{Text-to-Image dataset for billion-scale similarity search}.
\newblock
\urldef\tempurl%
\url{https://research.yandex.com/datasets/text-to-image-dataset-for-billion-scale-similarity-search}
\showURL{%
Retrieved April 13, 2025 from \tempurl}


\bibitem[Dobson et~al\mbox{.}(2023)]%
        {largescale}
\bibfield{author}{\bibinfo{person}{Magdalen Dobson}, \bibinfo{person}{Zheqi Shen}, \bibinfo{person}{Guy~E. Blelloch}, \bibinfo{person}{Laxman Dhulipala}, \bibinfo{person}{Yan Gu}, \bibinfo{person}{Harsha~Vardhan Simhadri}, {and} \bibinfo{person}{Yihan Sun}.} \bibinfo{year}{2023}\natexlab{}.
\newblock \showarticletitle{Scaling Graph-Based {ANNS} Algorithms to Billion-Size Datasets: {A} Comparative Analysis}.
\newblock \bibinfo{journal}{\emph{CoRR}}  \bibinfo{volume}{abs/2305.04359} (\bibinfo{year}{2023}).
\newblock
\showeprint[arXiv]{2305.04359}
\href{https://doi.org/10.48550/ARXIV.2305.04359}{doi:\nolinkurl{10.48550/ARXIV.2305.04359}}


\bibitem[Fu et~al\mbox{.}(2022)]%
        {ssg}
\bibfield{author}{\bibinfo{person}{Cong Fu}, \bibinfo{person}{Changxu Wang}, {and} \bibinfo{person}{Deng Cai}.} \bibinfo{year}{2022}\natexlab{}.
\newblock \showarticletitle{High Dimensional Similarity Search With Satellite System Graph: Efficiency, Scalability, and Unindexed Query Compatibility}.
\newblock \bibinfo{journal}{\emph{{IEEE} Trans. Pattern Anal. Mach. Intell.}} \bibinfo{volume}{44}, \bibinfo{number}{8} (\bibinfo{year}{2022}), \bibinfo{pages}{4139--4150}.
\newblock
\href{https://doi.org/10.1109/TPAMI.2021.3067706}{doi:\nolinkurl{10.1109/TPAMI.2021.3067706}}


\bibitem[Fu et~al\mbox{.}(2019)]%
        {nsg}
\bibfield{author}{\bibinfo{person}{Cong Fu}, \bibinfo{person}{Chao Xiang}, \bibinfo{person}{Changxu Wang}, {and} \bibinfo{person}{Deng Cai}.} \bibinfo{year}{2019}\natexlab{}.
\newblock \showarticletitle{Fast Approximate Nearest Neighbor Search With The Navigating Spreading-out Graph}.
\newblock \bibinfo{journal}{\emph{Proc. {VLDB} Endow.}} \bibinfo{volume}{12}, \bibinfo{number}{5} (\bibinfo{year}{2019}), \bibinfo{pages}{461--474}.
\newblock
\href{https://doi.org/10.14778/3303753.3303754}{doi:\nolinkurl{10.14778/3303753.3303754}}


\bibitem[Gionis et~al\mbox{.}(1999)]%
        {lsh}
\bibfield{author}{\bibinfo{person}{Aristides Gionis}, \bibinfo{person}{Piotr Indyk}, {and} \bibinfo{person}{Rajeev Motwani}.} \bibinfo{year}{1999}\natexlab{}.
\newblock \showarticletitle{Similarity Search in High Dimensions via Hashing}. In \bibinfo{booktitle}{\emph{VLDB'99, Proceedings of 25th International Conference on Very Large Data Bases, September 7-10, 1999, Edinburgh, Scotland, {UK}}}, \bibfield{editor}{\bibinfo{person}{Malcolm~P. Atkinson}, \bibinfo{person}{Maria~E. Orlowska}, \bibinfo{person}{Patrick Valduriez}, \bibinfo{person}{Stanley~B. Zdonik}, {and} \bibinfo{person}{Michael~L. Brodie}} (Eds.). \bibinfo{publisher}{Morgan Kaufmann}, \bibinfo{pages}{518--529}.
\newblock
\urldef\tempurl%
\url{http://www.vldb.org/conf/1999/P49.pdf}
\showURL{%
\tempurl}


\bibitem[Guo et~al\mbox{.}(2020)]%
        {avq}
\bibfield{author}{\bibinfo{person}{Ruiqi Guo}, \bibinfo{person}{Philip Sun}, \bibinfo{person}{Erik Lindgren}, \bibinfo{person}{Quan Geng}, \bibinfo{person}{David Simcha}, \bibinfo{person}{Felix Chern}, {and} \bibinfo{person}{Sanjiv Kumar}.} \bibinfo{year}{2020}\natexlab{}.
\newblock \showarticletitle{Accelerating Large-Scale Inference with Anisotropic Vector Quantization}. In \bibinfo{booktitle}{\emph{Proceedings of the 37th International Conference on Machine Learning, {ICML} 2020, 13-18 July 2020, Virtual Event}} \emph{(\bibinfo{series}{Proceedings of Machine Learning Research}, Vol.~\bibinfo{volume}{119})}. \bibinfo{publisher}{{PMLR}}, \bibinfo{pages}{3887--3896}.
\newblock
\urldef\tempurl%
\url{http://proceedings.mlr.press/v119/guo20h.html}
\showURL{%
\tempurl}


\bibitem[Harris et~al\mbox{.}(2025)]%
        {medicine1}
\bibfield{author}{\bibinfo{person}{Lee Harris}, \bibinfo{person}{Philippe~De Wilde}, {and} \bibinfo{person}{James Bentham}.} \bibinfo{year}{2025}\natexlab{}.
\newblock \showarticletitle{Comparing Lexical and Semantic Vector Search Methods When Classifying Medical Documents}.
\newblock \bibinfo{journal}{\emph{CoRR}}  \bibinfo{volume}{abs/2505.11582} (\bibinfo{year}{2025}).
\newblock
\showeprint[arXiv]{2505.11582}
\href{https://doi.org/10.48550/ARXIV.2505.11582}{doi:\nolinkurl{10.48550/ARXIV.2505.11582}}


\bibitem[{HiSilicon}(2024)]%
        {HiSilicon}
\bibfield{author}{\bibinfo{person}{{HiSilicon}}.} \bibinfo{year}{2024}\natexlab{}.
\newblock \bibinfo{title}{Kunpeng 920 Chipset}.
\newblock \bibinfo{howpublished}{\url{https://www.hisilicon.com/en/products/kunpeng/huawei-kunpeng/huawei-kunpeng-920}}.
\newblock
\newblock
\shownote{Accessed: 2025-07-29}.


\bibitem[Huang et~al\mbox{.}(2020)]%
        {fbembedding}
\bibfield{author}{\bibinfo{person}{Jui-Ting Huang}, \bibinfo{person}{Ashish Sharma}, \bibinfo{person}{Shuying Sun}, \bibinfo{person}{Li Xia}, \bibinfo{person}{David Zhang}, \bibinfo{person}{Philip Pronin}, \bibinfo{person}{Janani Padmanabhan}, \bibinfo{person}{Giuseppe Ottaviano}, {and} \bibinfo{person}{Linjun Yang}.} \bibinfo{year}{2020}\natexlab{}.
\newblock \showarticletitle{Embedding-based retrieval in facebook search}. In \bibinfo{booktitle}{\emph{Proceedings of the 26th ACM SIGKDD International Conference on Knowledge Discovery \& Data Mining}}. \bibinfo{pages}{2553--2561}.
\newblock


\bibitem[{Huawei Technologies Co., Ltd.}(2023)]%
        {hikunpeng2023}
\bibfield{author}{\bibinfo{person}{{Huawei Technologies Co., Ltd.}}} \bibinfo{year}{2023}\natexlab{}.
\newblock \bibinfo{title}{Kunpeng Computing}.
\newblock \bibinfo{howpublished}{\url{https://www.hikunpeng.com/zh}}.
\newblock
\newblock
\shownote{Accessed: 2023-12-01}.


\bibitem[Iacovides et~al\mbox{.}(2025)]%
        {finance1}
\bibfield{author}{\bibinfo{person}{Giorgos Iacovides}, \bibinfo{person}{Wuyang Zhou}, {and} \bibinfo{person}{Danilo Mandic}.} \bibinfo{year}{2025}\natexlab{}.
\newblock \bibinfo{title}{FinDPO: Financial Sentiment Analysis for Algorithmic Trading through Preference Optimization of LLMs}.
\newblock
\showeprint[arxiv]{2507.18417}~[cs.CL]
\urldef\tempurl%
\url{https://arxiv.org/abs/2507.18417}
\showURL{%
\tempurl}


\bibitem[Indyk and Motwani(1998)]%
        {approximate_graph}
\bibfield{author}{\bibinfo{person}{Piotr Indyk} {and} \bibinfo{person}{Rajeev Motwani}.} \bibinfo{year}{1998}\natexlab{}.
\newblock \showarticletitle{Approximate nearest neighbors: towards removing the curse of dimensionality}. In \bibinfo{booktitle}{\emph{Proceedings of the Thirtieth Annual ACM Symposium on Theory of Computing}} (Dallas, Texas, USA) \emph{(\bibinfo{series}{STOC '98})}. \bibinfo{publisher}{Association for Computing Machinery}, \bibinfo{address}{New York, NY, USA}, \bibinfo{pages}{604–613}.
\newblock
\showISBNx{0897919629}
\href{https://doi.org/10.1145/276698.276876}{doi:\nolinkurl{10.1145/276698.276876}}


\bibitem[J{\'{e}}gou et~al\mbox{.}(2011)]%
        {pq}
\bibfield{author}{\bibinfo{person}{Herv{\'{e}} J{\'{e}}gou}, \bibinfo{person}{Matthijs Douze}, {and} \bibinfo{person}{Cordelia Schmid}.} \bibinfo{year}{2011}\natexlab{}.
\newblock \showarticletitle{Product Quantization for Nearest Neighbor Search}.
\newblock \bibinfo{journal}{\emph{{IEEE} Trans. Pattern Anal. Mach. Intell.}} \bibinfo{volume}{33}, \bibinfo{number}{1} (\bibinfo{year}{2011}), \bibinfo{pages}{117--128}.
\newblock
\href{https://doi.org/10.1109/TPAMI.2010.57}{doi:\nolinkurl{10.1109/TPAMI.2010.57}}


\bibitem[Johnson et~al\mbox{.}(2021a)]%
        {ivfpq}
\bibfield{author}{\bibinfo{person}{Jeff Johnson}, \bibinfo{person}{Matthijs Douze}, {and} \bibinfo{person}{Herv{\'{e}} J{\'{e}}gou}.} \bibinfo{year}{2021}\natexlab{a}.
\newblock \showarticletitle{Billion-Scale Similarity Search with GPUs}.
\newblock \bibinfo{journal}{\emph{{IEEE} Trans. Big Data}} \bibinfo{volume}{7}, \bibinfo{number}{3} (\bibinfo{year}{2021}), \bibinfo{pages}{535--547}.
\newblock
\href{https://doi.org/10.1109/TBDATA.2019.2921572}{doi:\nolinkurl{10.1109/TBDATA.2019.2921572}}


\bibitem[Johnson et~al\mbox{.}(2021b)]%
        {faiss}
\bibfield{author}{\bibinfo{person}{Jeff Johnson}, \bibinfo{person}{Matthijs Douze}, {and} \bibinfo{person}{Hervé Jégou}.} \bibinfo{year}{2021}\natexlab{b}.
\newblock \showarticletitle{Billion-Scale Similarity Search with GPUs}.
\newblock \bibinfo{journal}{\emph{IEEE Transactions on Big Data}} \bibinfo{volume}{7}, \bibinfo{number}{3} (\bibinfo{year}{2021}), \bibinfo{pages}{535--547}.
\newblock
\href{https://doi.org/10.1109/TBDATA.2019.2921572}{doi:\nolinkurl{10.1109/TBDATA.2019.2921572}}


\bibitem[Kalantidis and Avrithis(2014)]%
        {localpq}
\bibfield{author}{\bibinfo{person}{Yannis Kalantidis} {and} \bibinfo{person}{Yannis Avrithis}.} \bibinfo{year}{2014}\natexlab{}.
\newblock \showarticletitle{Locally optimized product quantization for approximate nearest neighbor search}. In \bibinfo{booktitle}{\emph{Proceedings of the IEEE conference on computer vision and pattern recognition}}. \bibinfo{pages}{2321--2328}.
\newblock


\bibitem[Laurent~Amsaleg(2010)]%
        {sift}
\bibfield{author}{\bibinfo{person}{Hervé~Jégou Laurent~Amsaleg}.} \bibinfo{year}{2010}\natexlab{}.
\newblock \bibinfo{title}{Datasets for approximate nearest neighbor search}.
\newblock
\urldef\tempurl%
\url{http://corpus-texmex.irisa.fr/}
\showURL{%
\tempurl}
\newblock
\shownote{Accessed: 2025-04-17}.


\bibitem[Lee et~al\mbox{.}(2022)]%
        {anna}
\bibfield{author}{\bibinfo{person}{Yejin Lee}, \bibinfo{person}{Hyunji Choi}, \bibinfo{person}{Sunhong Min}, \bibinfo{person}{Hyunseung Lee}, \bibinfo{person}{Sangwon Beak}, \bibinfo{person}{Dawoon Jeong}, \bibinfo{person}{Jae~W. Lee}, {and} \bibinfo{person}{Tae~Jun Ham}.} \bibinfo{year}{2022}\natexlab{}.
\newblock \showarticletitle{ANNA: Specialized Architecture for Approximate Nearest Neighbor Search}. In \bibinfo{booktitle}{\emph{2022 IEEE International Symposium on High-Performance Computer Architecture (HPCA)}}. \bibinfo{pages}{169--183}.
\newblock
\href{https://doi.org/10.1109/HPCA53966.2022.00021}{doi:\nolinkurl{10.1109/HPCA53966.2022.00021}}


\bibitem[Li et~al\mbox{.}(2019)]%
        {anns}
\bibfield{author}{\bibinfo{person}{Wen Li}, \bibinfo{person}{Ying Zhang}, \bibinfo{person}{Yifang Sun}, \bibinfo{person}{Wei Wang}, \bibinfo{person}{Mingjie Li}, \bibinfo{person}{Wenjie Zhang}, {and} \bibinfo{person}{Xuemin Lin}.} \bibinfo{year}{2019}\natexlab{}.
\newblock \showarticletitle{Approximate nearest neighbor search on high dimensional data—experiments, analyses, and improvement}.
\newblock \bibinfo{journal}{\emph{IEEE Transactions on Knowledge and Data Engineering}} \bibinfo{volume}{32}, \bibinfo{number}{8} (\bibinfo{year}{2019}), \bibinfo{pages}{1475--1488}.
\newblock


\bibitem[Li et~al\mbox{.}(2020)]%
        {vectorsearchsurvey}
\bibfield{author}{\bibinfo{person}{Wen Li}, \bibinfo{person}{Ying Zhang}, \bibinfo{person}{Yifang Sun}, \bibinfo{person}{Wei Wang}, \bibinfo{person}{Mingjie Li}, \bibinfo{person}{Wenjie Zhang}, {and} \bibinfo{person}{Xuemin Lin}.} \bibinfo{year}{2020}\natexlab{}.
\newblock \showarticletitle{Approximate Nearest Neighbor Search on High Dimensional Data - Experiments, Analyses, and Improvement}.
\newblock \bibinfo{journal}{\emph{{IEEE} Trans. Knowl. Data Eng.}} \bibinfo{volume}{32}, \bibinfo{number}{8} (\bibinfo{year}{2020}), \bibinfo{pages}{1475--1488}.
\newblock
\href{https://doi.org/10.1109/TKDE.2019.2909204}{doi:\nolinkurl{10.1109/TKDE.2019.2909204}}


\bibitem[Liu et~al\mbox{.}(2025)]%
        {llmbot2}
\bibfield{author}{\bibinfo{person}{Shige Liu}, \bibinfo{person}{Zhifang Zeng}, \bibinfo{person}{Li Chen}, \bibinfo{person}{Adil Ainihaer}, \bibinfo{person}{Arun Ramasami}, \bibinfo{person}{Songting Chen}, \bibinfo{person}{Yu Xu}, \bibinfo{person}{Mingxi Wu}, {and} \bibinfo{person}{Jianguo Wang}.} \bibinfo{year}{2025}\natexlab{}.
\newblock \bibinfo{title}{TigerVector: Supporting Vector Search in Graph Databases for Advanced RAGs}.
\newblock
\showeprint[arxiv]{2501.11216}~[cs.DB]
\urldef\tempurl%
\url{https://arxiv.org/abs/2501.11216}
\showURL{%
\tempurl}


\bibitem[Malkov and Yashunin(2018)]%
        {hnswlib}
\bibfield{author}{\bibinfo{person}{Yu.~A. Malkov} {and} \bibinfo{person}{D.~A. Yashunin}.} \bibinfo{year}{2018}\natexlab{}.
\newblock \bibinfo{title}{hnswlib: Header-only C++ library for fast approximate nearest neighbors search}.
\newblock \bibinfo{howpublished}{\url{https://github.com/nmslib/hnswlib}}.
\newblock


\bibitem[Malkov and Yashunin(2020)]%
        {hnsw}
\bibfield{author}{\bibinfo{person}{Yury~A. Malkov} {and} \bibinfo{person}{Dmitry~A. Yashunin}.} \bibinfo{year}{2020}\natexlab{}.
\newblock \showarticletitle{Efficient and Robust Approximate Nearest Neighbor Search Using Hierarchical Navigable Small World Graphs}.
\newblock \bibinfo{journal}{\emph{{IEEE} Trans. Pattern Anal. Mach. Intell.}} \bibinfo{volume}{42}, \bibinfo{number}{4} (\bibinfo{year}{2020}), \bibinfo{pages}{824--836}.
\newblock
\href{https://doi.org/10.1109/TPAMI.2018.2889473}{doi:\nolinkurl{10.1109/TPAMI.2018.2889473}}


\bibitem[Ono and Matsui(2023)]%
        {rnndescent}
\bibfield{author}{\bibinfo{person}{Naoki Ono} {and} \bibinfo{person}{Yusuke Matsui}.} \bibinfo{year}{2023}\natexlab{}.
\newblock \showarticletitle{Relative NN-Descent: {A} Fast Index Construction for Graph-Based Approximate Nearest Neighbor Search}. In \bibinfo{booktitle}{\emph{Proceedings of the 31st {ACM} International Conference on Multimedia, {MM} 2023, Ottawa, ON, Canada, 29 October 2023- 3 November 2023}}, \bibfield{editor}{\bibinfo{person}{Abdulmotaleb El{-}Saddik}, \bibinfo{person}{Tao Mei}, \bibinfo{person}{Rita Cucchiara}, \bibinfo{person}{Marco Bertini}, \bibinfo{person}{Diana Patricia~Tobon Vallejo}, \bibinfo{person}{Pradeep~K. Atrey}, {and} \bibinfo{person}{M.~Shamim Hossain}} (Eds.). \bibinfo{publisher}{{ACM}}, \bibinfo{pages}{1659--1667}.
\newblock
\href{https://doi.org/10.1145/3581783.3612290}{doi:\nolinkurl{10.1145/3581783.3612290}}


\bibitem[{openGauss Community}(2023)]%
        {opengauss}
\bibfield{author}{\bibinfo{person}{{openGauss Community}}.} \bibinfo{year}{2023}\natexlab{}.
\newblock \bibinfo{title}{openGauss: An Open-Source Enterprise-Class Relational Database}.
\newblock \bibinfo{howpublished}{Official Website}.
\newblock
\urldef\tempurl%
\url{https://opengauss.org/en/}
\showURL{%
\tempurl}
\newblock
\shownote{An open-source RDBMS supported by Huawei}.


\bibitem[Pennington et~al\mbox{.}(2014)]%
        {glove}
\bibfield{author}{\bibinfo{person}{Jeffrey Pennington}, \bibinfo{person}{Richard Socher}, {and} \bibinfo{person}{Christopher~D. Manning}.} \bibinfo{year}{2014}\natexlab{}.
\newblock \bibinfo{title}{{GloVe}: Global Vectors for Word Representation}.
\newblock \bibinfo{howpublished}{\url{https://nlp.stanford.edu/projects/glove/}}.
\newblock
\newblock
\shownote{Accessed: 2025-07-31}.


\bibitem[Radford et~al\mbox{.}(2021)]%
        {learning}
\bibfield{author}{\bibinfo{person}{Alec Radford}, \bibinfo{person}{Jong~Wook Kim}, \bibinfo{person}{Chris Hallacy}, \bibinfo{person}{Aditya Ramesh}, \bibinfo{person}{Gabriel Goh}, \bibinfo{person}{Sandhini Agarwal}, \bibinfo{person}{Girish Sastry}, \bibinfo{person}{Amanda Askell}, \bibinfo{person}{Pamela Mishkin}, \bibinfo{person}{Jack Clark}, {et~al\mbox{.}}} \bibinfo{year}{2021}\natexlab{}.
\newblock \showarticletitle{Learning transferable visual models from natural language supervision}. In \bibinfo{booktitle}{\emph{International conference on machine learning}}. PmLR, \bibinfo{pages}{8748--8763}.
\newblock


\bibitem[Research(2020)]%
        {scann}
\bibfield{author}{\bibinfo{person}{Google Research}.} \bibinfo{year}{2020}\natexlab{}.
\newblock \bibinfo{booktitle}{\emph{ScaNN: Efficient Vector Similarity Search}}.
\newblock


\bibitem[Sarmah et~al\mbox{.}(2024)]%
        {llmbot1}
\bibfield{author}{\bibinfo{person}{Bhaskarjit Sarmah}, \bibinfo{person}{Benika Hall}, \bibinfo{person}{Rohan Rao}, \bibinfo{person}{Sunil Patel}, \bibinfo{person}{Stefano Pasquali}, {and} \bibinfo{person}{Dhagash Mehta}.} \bibinfo{year}{2024}\natexlab{}.
\newblock \bibinfo{title}{HybridRAG: Integrating Knowledge Graphs and Vector Retrieval Augmented Generation for Efficient Information Extraction}.
\newblock
\showeprint[arxiv]{2408.04948}~[cs.CL]
\urldef\tempurl%
\url{https://arxiv.org/abs/2408.04948}
\showURL{%
\tempurl}


\bibitem[Scott(1992)]%
        {dimcurse}
\bibfield{author}{\bibinfo{person}{David~W. Scott}.} \bibinfo{year}{1992}\natexlab{}.
\newblock \bibinfo{booktitle}{\emph{Multivariate Density Estimation: Theory, Practice, and Visualization}}.
\newblock \bibinfo{publisher}{Wiley}.
\newblock
\showISBNx{978-0-47154770-9}
\href{https://doi.org/10.1002/9780470316849}{doi:\nolinkurl{10.1002/9780470316849}}


\bibitem[Shi et~al\mbox{.}(2022)]%
        {medicine2}
\bibfield{author}{\bibinfo{person}{Wen Shi}, \bibinfo{person}{Jianling Liu}, \bibinfo{person}{Jingyu Zhang}, \bibinfo{person}{Yuran Men}, \bibinfo{person}{Hongwei Chen}, \bibinfo{person}{Deke Wang}, {and} \bibinfo{person}{Yang Cao}.} \bibinfo{year}{2022}\natexlab{}.
\newblock \showarticletitle{Feature Selection and Parameter Optimization of Support Vector Machines Based on a Local Search Based Firefly Algorithm for Classification of Formulas in Traditional Chinese Medicine}.
\newblock \bibinfo{journal}{\emph{{IEICE} Trans. Fundam. Electron. Commun. Comput. Sci.}} \bibinfo{volume}{105-A}, \bibinfo{number}{5} (\bibinfo{year}{2022}), \bibinfo{pages}{882--886}.
\newblock
\href{https://doi.org/10.1587/TRANSFUN.2021EAL2075}{doi:\nolinkurl{10.1587/TRANSFUN.2021EAL2075}}


\bibitem[Silpa{-}Anan and Hartley(2008)]%
        {kdtree}
\bibfield{author}{\bibinfo{person}{Chanop Silpa{-}Anan} {and} \bibinfo{person}{Richard~I. Hartley}.} \bibinfo{year}{2008}\natexlab{}.
\newblock \showarticletitle{Optimised KD-trees for fast image descriptor matching}. In \bibinfo{booktitle}{\emph{2008 {IEEE} Computer Society Conference on Computer Vision and Pattern Recognition {(CVPR} 2008), 24-26 June 2008, Anchorage, Alaska, {USA}}}. \bibinfo{publisher}{{IEEE} Computer Society}.
\newblock
\href{https://doi.org/10.1109/CVPR.2008.4587638}{doi:\nolinkurl{10.1109/CVPR.2008.4587638}}


\bibitem[Subramanya et~al\mbox{.}(2019)]%
        {diskann}
\bibfield{author}{\bibinfo{person}{Suhas~Jayaram Subramanya}, \bibinfo{person}{Devvrit}, \bibinfo{person}{Rohan Kadekodi}, \bibinfo{person}{Ravishankar Krishaswamy}, {and} \bibinfo{person}{Harsha~Vardhan Simhadri}.} \bibinfo{year}{2019}\natexlab{}.
\newblock \bibinfo{booktitle}{\emph{DiskANN: fast accurate billion-point nearest neighbor search on a single node}}.
\newblock \bibinfo{publisher}{Curran Associates Inc.}, \bibinfo{address}{Red Hook, NY, USA}.
\newblock


\bibitem[Sun et~al\mbox{.}(2023)]%
        {soar}
\bibfield{author}{\bibinfo{person}{Philip Sun}, \bibinfo{person}{David Simcha}, \bibinfo{person}{Dave Dopson}, \bibinfo{person}{Ruiqi Guo}, {and} \bibinfo{person}{Sanjiv Kumar}.} \bibinfo{year}{2023}\natexlab{}.
\newblock \showarticletitle{{SOAR:} Improved Indexing for Approximate Nearest Neighbor Search}. In \bibinfo{booktitle}{\emph{Advances in Neural Information Processing Systems 36: Annual Conference on Neural Information Processing Systems 2023, NeurIPS 2023, New Orleans, LA, USA, December 10 - 16, 2023}}, \bibfield{editor}{\bibinfo{person}{Alice Oh}, \bibinfo{person}{Tristan Naumann}, \bibinfo{person}{Amir Globerson}, \bibinfo{person}{Kate Saenko}, \bibinfo{person}{Moritz Hardt}, {and} \bibinfo{person}{Sergey Levine}} (Eds.).
\newblock
\urldef\tempurl%
\url{http://papers.nips.cc/paper\_files/paper/2023/hash/0973524e02a712af33325d0688ae6f49-Abstract-Conference.html}
\showURL{%
\tempurl}


\bibitem[Van~Gysel et~al\mbox{.}(2016)]%
        {van2016learning}
\bibfield{author}{\bibinfo{person}{Christophe Van~Gysel}, \bibinfo{person}{Maarten de Rijke}, {and} \bibinfo{person}{Evangelos Kanoulas}.} \bibinfo{year}{2016}\natexlab{}.
\newblock \showarticletitle{Learning latent vector spaces for product search}. In \bibinfo{booktitle}{\emph{Proceedings of the 25th ACM international on conference on information and knowledge management}}. \bibinfo{pages}{165--174}.
\newblock


\bibitem[Wang et~al\mbox{.}(2021)]%
        {milvus}
\bibfield{author}{\bibinfo{person}{Jianguo Wang}, \bibinfo{person}{Xiaomeng Yi}, \bibinfo{person}{Rentong Guo}, \bibinfo{person}{Hai Jin}, \bibinfo{person}{Peng Xu}, \bibinfo{person}{Shengjun Li}, \bibinfo{person}{Xiangyu Wang}, \bibinfo{person}{Xiangzhou Guo}, \bibinfo{person}{Chengming Li}, \bibinfo{person}{Xiaohai Xu}, \bibinfo{person}{Kun Yu}, \bibinfo{person}{Yuxing Yuan}, \bibinfo{person}{Yinghao Zou}, \bibinfo{person}{Jiquan Long}, \bibinfo{person}{Yudong Cai}, \bibinfo{person}{Zhenxiang Li}, \bibinfo{person}{Zhifeng Zhang}, \bibinfo{person}{Yihua Mo}, \bibinfo{person}{Jun Gu}, \bibinfo{person}{Ruiyi Jiang}, \bibinfo{person}{Yi Wei}, {and} \bibinfo{person}{Charles Xie}.} \bibinfo{year}{2021}\natexlab{}.
\newblock \showarticletitle{Milvus: A Purpose-Built Vector Data Management System}. In \bibinfo{booktitle}{\emph{Proceedings of the 2021 International Conference on Management of Data}} (Virtual Event, China) \emph{(\bibinfo{series}{SIGMOD '21})}. \bibinfo{publisher}{Association for Computing Machinery}, \bibinfo{address}{New York, NY, USA}, \bibinfo{pages}{2614–2627}.
\newblock
\showISBNx{9781450383431}
\href{https://doi.org/10.1145/3448016.3457550}{doi:\nolinkurl{10.1145/3448016.3457550}}


\bibitem[Wang(2025)]%
        {PyGlass}
\bibfield{author}{\bibinfo{person}{Zihao Wang}.} \bibinfo{year}{2025}\natexlab{}.
\newblock \bibinfo{title}{Graph Library for Approximate Similarity Search}.
\newblock
\urldef\tempurl%
\url{https://github.com/zilliztech/pyglass}
\showURL{%
\tempurl}


\bibitem[Wei et~al\mbox{.}(2016)]%
        {gorder}
\bibfield{author}{\bibinfo{person}{Hao Wei}, \bibinfo{person}{Jeffrey~Xu Yu}, \bibinfo{person}{Can Lu}, {and} \bibinfo{person}{Xuemin Lin}.} \bibinfo{year}{2016}\natexlab{}.
\newblock \showarticletitle{Speedup Graph Processing by Graph Ordering}. In \bibinfo{booktitle}{\emph{Proceedings of the 2016 International Conference on Management of Data}} (San Francisco, California, USA) \emph{(\bibinfo{series}{SIGMOD '16})}. \bibinfo{publisher}{Association for Computing Machinery}, \bibinfo{address}{New York, NY, USA}, \bibinfo{pages}{1813–1828}.
\newblock
\showISBNx{9781450335317}
\href{https://doi.org/10.1145/2882903.2915220}{doi:\nolinkurl{10.1145/2882903.2915220}}


\bibitem[Weiss et~al\mbox{.}(2008)]%
        {spectral_hashing}
\bibfield{author}{\bibinfo{person}{Yair Weiss}, \bibinfo{person}{Antonio Torralba}, {and} \bibinfo{person}{Robert Fergus}.} \bibinfo{year}{2008}\natexlab{}.
\newblock \showarticletitle{Spectral Hashing}. In \bibinfo{booktitle}{\emph{Advances in Neural Information Processing Systems 21, Proceedings of the Twenty-Second Annual Conference on Neural Information Processing Systems, Vancouver, British Columbia, Canada, December 8-11, 2008}}, \bibfield{editor}{\bibinfo{person}{Daphne Koller}, \bibinfo{person}{Dale Schuurmans}, \bibinfo{person}{Yoshua Bengio}, {and} \bibinfo{person}{L{\'{e}}on Bottou}} (Eds.). \bibinfo{publisher}{Curran Associates, Inc.}, \bibinfo{pages}{1753--1760}.
\newblock
\urldef\tempurl%
\url{https://proceedings.neurips.cc/paper/2008/hash/d58072be2820e8682c0a27c0518e805e-Abstract.html}
\showURL{%
\tempurl}


\bibitem[Zhang et~al\mbox{.}(2018)]%
        {visual}
\bibfield{author}{\bibinfo{person}{Yanhao Zhang}, \bibinfo{person}{Pan Pan}, \bibinfo{person}{Yun Zheng}, \bibinfo{person}{Kang Zhao}, \bibinfo{person}{Yingya Zhang}, \bibinfo{person}{Xiaofeng Ren}, {and} \bibinfo{person}{Rong Jin}.} \bibinfo{year}{2018}\natexlab{}.
\newblock \showarticletitle{Visual search at alibaba}. In \bibinfo{booktitle}{\emph{Proceedings of the 24th ACM SIGKDD international conference on knowledge discovery \& data mining}}. \bibinfo{pages}{993--1001}.
\newblock


\bibitem[Zhao et~al\mbox{.}(2021)]%
        {zhaolearning}
\bibfield{author}{\bibinfo{person}{Yifan Zhao}, \bibinfo{person}{Huiyu Cai}, \bibinfo{person}{Zuobai Zhang}, \bibinfo{person}{Jian Tang}, {and} \bibinfo{person}{Yue Li}.} \bibinfo{year}{2021}\natexlab{}.
\newblock \showarticletitle{Learning interpretable cellular and gene signature embeddings from single-cell transcriptomic data}.
\newblock \bibinfo{journal}{\emph{Nature communications}} \bibinfo{volume}{12}, \bibinfo{number}{1} (\bibinfo{year}{2021}), \bibinfo{pages}{5261}.
\newblock


\end{thebibliography}

\end{document}